\documentclass[11pt, ]{article}
\usepackage{epsfig,amsmath,amssymb,euscript}
\renewcommand{\baselinestretch}{1.7}
\makeatletter

 \@addtoreset{equation}{section}
 
\newtheorem{theorem}{Theorem}
\newtheorem{lemma}{Lemma} 
\newtheorem{corollary}{Corollary}

\renewcommand{\hat}{\widehat}
\def\singlespace{\def\baselinestretch{1}\@normalsize}

\newcommand{\bA}{{\mathbf A}}
\newcommand{\bB}{{\mathbf B}}

\newcommand{\bI}{{\mathbf I}}

\newcommand{\bV}{{\mathbf V}}
\newcommand{\bW}{{\mathbf W}}

\newcommand{\bY}{{\mathbf Y}}

\newcommand{\ba}{{\mathbf a}}

\newcommand{\be}{{\mathbf e}}

\newcommand{\bm}{{\mathbf m}}

\newcommand{\bu}{{\mathbf u}}
\newcommand{\bv}{{\mathbf v}}
\newcommand{\bw}{{\mathbf w}}
\newcommand{\bx}{{\mathbf x}}
\newcommand{\by}{{\mathbf y}}
\newcommand{\bz}{{\mathbf z}}

\newcommand{\blambda}{\boldsymbol{\lambda}}

\newcommand{\bSigma}{\boldsymbol{\Sigma}}

\newcommand{\bve}{\mbox{\boldmath$\varepsilon$}}

\newcommand{\bmu} {\boldsymbol{\mu}}

\newcommand{\bD}{{\mathbf D}}

\def\6bullets{\bullet\bullet\bullet\bullet\bullet\bullet}

\newcommand{\qqed}{\hfill $\Box$ \vspace{1mm}}

\begin{document}
\title{\bf Modelling high-dimensional time series efficiently by means of constrained spatio--temporal models}

\author{Maria Lucia Parrella\footnote{Maria Lucia Parrella, Universit\`{a} di Salerno - Via Giovanni Paolo II 137 - 84084 Fisciano (SA), Italy, tel: +3908996221, fax: +39089962049, email: mparrella@unisa.it} \\
{\sl Department of Economics and Statistics, University of Salerno, Italy} \\
mparrella@unisa.it
}

\maketitle
\begin{abstract}
Many econometric analyses involve spatio--temporal data, where the concept of space may be interpreted in a wide sense (from a physical, economic, or social point of view).
A considerable amount of literature has addressed spatio--temporal models, with \emph{Spatial Dynamic Panel Data} (\emph{SDPD}) being widely investigated and applied.
In real data applications, checking the validity of the theoretical assumptions underlying the \emph{SDPD} models is essential but sometimes difficult. At other times, the assumptions are clearly violated. For example, the \emph{spatial matrix} (which gives the spatial weights and influences the correlation structure of the process) is assumed to be known, but it may actually be unknown and needs to be estimated.
In such cases, the properties and performance of the \emph{SDPD} model's estimator are generally affected.
Motivated by such considerations, we propose a new model (called \emph{stationary SDPD}) and a new estimation procedure based on very simple and clear assumptions that can be easily checked using real data. The new model is highly adaptive, and the estimation procedure has a rate of convergence that is not affected by the dimension of the time series (under general assumptions), notwithstanding the relatively high number of parameters to be estimated. A further result of this work is that the new model may also be used to represent a wide class of multivariate time series that are not explicitly spatio--temporal, with respect to a ``latent spatial matrix,'' which needs to be estimated. Therefore, it can be used as a valid alternative to vector autoregressive (\emph{VAR}) models with two immediate advantages: \emph{i}) a faster rate of convergence of the estimation procedure and \emph{ii}) the possibility of estimating the model even when the dimension is higher than the time series length, overcoming the \emph{curse of dimensionality} typical of the \emph{VAR} models. The simulation study shows that the new estimation procedure performs well compared with the classic alternative procedure, even when the spatial matrix is unknown, and therefore, to be estimated.
\end{abstract}

\section{Introduction}
Spatio--temporal models are widely used by practitioners. Explaining economic, environmental, social, or biological phenomena, such as peer influence, neighbourhood effects, contagion, epidemics, interdependent preferences, climate change, and so on, are only some of the interesting applications of such models. A widely used spatio--temporal model is the spatial dynamic panel data model (SDPD) proposed and analysed by \cite{LeeYu10a}. See \cite{LeeYu10b} for a survey. To improve adaptivity of SDPD models, \cite{DouAlt15} recently proposed a generalized model that assigns different coefficients to varied locations and assumes heteroskedastic and spatially correlated errors. The model is
\begin{equation}\label{eqn1}
\by_t = D(\blambda_0)\bW\by_t + D({\blambda_1})\by_{t-1} + D(\blambda_2)\bW\by_{t-1} + \bve_t,
\end{equation}
where the vector $\by_t$ is of order $p$ and contains the observations at time $t$ from $p$ different locations; the errors $\bve_t$ are serially uncorrelated; the \emph{spatial matrix} $\bW$ is a weight matrix with zero main diagonal and is assumed to be known; $D(\blambda_j)$ with $j=0,1,2$ are diagonal matrices, and $\blambda_j$ are the vectors with the spatial coefficients $\lambda_{ji}$ for $i=1,\ldots,p$. The \emph{generalized SDPD} model in (\ref{eqn1}) guarantees adaptivity by means of its $3p$ parameters. It is characterized by the sum of three terms: the \emph{spatial component}, driven by matrix $\bW$ and the spatial parameter $\blambda_0$; the \emph{dynamic component}, driven by the autoregressive parameter $\blambda_1$; and the \emph{spatial--dynamic component}, driven by matrix $\bW$ and the spatial--autoregressive parameter $\blambda_2$. If the vectors $\blambda_j$ are scalars for all $j$, then model (\ref{eqn1}) reduces to the classic SDPD of \cite{LeeYu10a}.

The errors $\bve_t$ in model (\ref{eqn1}) are serially uncorrelated and may show heteroskedasticity and cross-correlation over space, so that $\mathop{var}(\bve_t)$ is a full matrix. This is a novelty compared with the \emph{SDPD} model of \cite{LeeYu10a}, where the errors must be cross-uncorrelated and homoskedastic in order to get consistency of the estimators. A setup similar to ours for the errors has been also considered by \cite{KelPru10} and \cite{Su12}, but not for panel models. However, their estimators are generally based on the instrumental variables technique, in order to overcome the endogeneity of the \emph{zero-lag} regressor. For the \emph{generalized SDPD} model, instead, \cite{DouAlt15} propose a new estimation procedure based on a generalized Yule--Walker approach. They show the consistency of the estimators under regularity assumptions. They also derive the convergence rate and the conditions under which the estimation procedure does not suffer for high-dimensional setups, notwithstanding the large number of parameters to be estimated (which become infinite with the dimension $p$).

In real data applications, it is important to check the validity of the assumptions required for the consistency of the estimation procedure. See, for example, the assumptions and asymptotic analysis in \cite{LeeYu10a} and \cite{DouAlt15} as well as the references therein. Checking such assumptions on real data is often not easy; at times, they are clearly violated.
Moreover, the spatial matrix $\bW$ is assumed to be known, but in many cases, this is not true, and it must be estimated. For example, the spatial weights can be associated with ``similarities'' between spatial units and measured by estimated correlations. Another example is when the spatial weights are zeroes/ones, depending on the ``relationships'' between the spatial units, but the neighbourhood structure of $\bW$ is unknown (\emph{i.e.}, it is not known where the ones must be allocated). In such cases, the performance of the \emph{SDPD} models needs to be investigated. Readers are advised to refer to recent papers on spatial matrix estimation (see, among others, \cite{LamSou16}).

Motivated by the above considerations, we propose a new version of the \emph{ SDPD} model obtained by adding a constraint to the spatial parameters of the \emph{generalized SDPD} of \cite{DouAlt15}. New estimators of the parameters are proposed and investigated theoretically and empirically.

The new model is called \emph{stationary SDPD} and has several advantages.
First, the structure of the model and the interpretation of the parameters are simpler than the \emph{generalized SDPD} model, with the consequence that the assumptions underlying the theoretical results are clearer and can be checked easily with real data. Moreover, the estimation procedure is fast and simple to implement.

Second, the proposed estimators of the parameters are always unbiased and reach the $\sqrt{T}$ convergence rate (where $T$ is the temporal length of the time series) even in the high-dimensional case, although the number of parameters tends to infinity with the dimension $p$.

Last, but not least, our model allows wider application than the classic \emph{SDPD} model, and it is general enough to represent a wide range of multivariate linear processes that can be implicitly interpreted (when they are not explicitly interpretable) as spatio--temporal processes, with respect to a ``latent spatial matrix,'' which needs to be estimated. A big implication of this is that our model is not necessarily confined to the representation of strict spatio--temporal processes (where the spatial matrix is known), but it can also be considered as a valid alternative to the general \emph{VAR} models (where there is no spatial matrix), with two relevant advantages: i) more efficient estimation of the model and ii) the possibility of estimating the model even when $p>T$, thus avoiding the \emph{curse of dimensionality} that characterizes the \emph{VAR} models. Surprisingly, the simulation results show the remarkably better performance of our model and the new estimation procedure compared with the standard VAR model and the standard estimation procedure, even when the spatial matrix is latent and, therefore, to be estimated (see section \ref{matrixA}).

The rest of the paper is organized as follows. Section \ref{sdpd} presents the new model and discusses the issue of identifiability. Section \ref{est_alg} describes the estimation procedure. The theoretical results are shown in section \ref{asymptotic}. The empirical performance of the estimation procedure is investigated in section \ref{simulazioni}. Finally, all the proofs are provided in the Appendix.

\section{A constrained spatio--temporal model: the stationary SDPD}\label{sdpd}
In the sequel, we assume that $\by_1, \cdots, \by_T$ are the observations from a stationary process defined by (\ref{eqn3}). The transpose of a matrix $\bA$ is denoted with $\bA^T$.
We assume that the process has mean zero and denote with $\bSigma_j=\mathop{cov}(\by_t,\by_{t-j})=E(\by_t\by_{t-j}^T)$ the covariance matrix of the process at the lag $j$.
The \emph{generalized SDPD} model in (\ref{eqn1}) can be reformulated as follows.
\begin{equation}
\left[\bI_p-D(\blambda_0)\bW\right]\by_t = D({\blambda_1})\left[\bI_p - D(\blambda^+_2)\bW\right]\by_{t-1} + \bve_t,  \label{eqn3}
\end{equation}
where $\blambda^+_2$  is a vector obtained by dividing the elements of $\blambda_2$ by the corresponding elements of $\blambda_1$ (assuming, for now, that all the coefficients in $\blambda_1$ are different from zero). Note that model (\ref{eqn3}) is equivalent to a multivariate (auto)regression between a linear combination of $\by_t$ and a linear combination of the lag $\by_{t-1}$, where the weights of the two linear combinations depend on $\bW$ and the coefficients $\blambda_0$ and $\blambda^+_2$, respectively.
\begin{equation}\label{zeta}
\bz_t^{(\blambda_0,\bW)} = D({\blambda_1})\bz_{t-1}^{(\blambda^+_2,\bW)} + \bve_t.
\end{equation}

Some special cases may arise from model (\ref{eqn3}) by adding restrictions on the parameters $\blambda_{j}$.
First, if we assume that the spatial parameters are constant over space, that is, $\blambda_{j}$ is scalar for $j=0,1,2$, then we obtain the classic \emph{SDPD} model of \cite{LeeYu10a}.

Another constrained model, proposed and analysed in this paper, may be derived by assuming that $\blambda_0=\blambda^+_2$.
The reason underlying the choice of this constraint is a generalized assumption of stationarity. In time series analysis, stationarity means that the dependence structure of the process is constant (in some sense) over time. In particular, second-order stationarity assumes that correlations between the observations $(\by_t,\by_{t-j})$ depend on the lag $j$ but not on $t$, implying that $\mathop{var}(\by_t)$ is constant for all $t$. However, in spatio--temporal time series, there are two kinds of correlations: \emph{temporal correlations}, involving observations at different time points, and \emph{spatial correlations}, involving observations at different spatial units. As we refer to stationarity, it makes sense to assume that spatial correlations are also time-invariant, which means that the weights in (\ref{zeta}) must not change over time, thus, $\left[\bI_p-D(\blambda_0)\bW\right]=\left[\bI_p-D(\blambda^+_2)\bW\right]$, also implying that $\mathop{var}(\bz_t)$ is the same for all $t$. Therefore, we add the constraint $\blambda_0=\blambda^+_2$, and the model becomes
\begin{equation} \label{b1}
\left[\bI_p-D(\blambda_0)\bW\right]\by_t = D(\blambda_1)\left[\bI_p-D(\blambda_0)\bW\right]\by_{t-1} +  \bve_t.
\end{equation}

We denote the model as \emph{stationary SDPD}.
Model (\ref{b1}) has several advantages that will be shown in the following sections.  Above all, imposing spatio--temporal stationarity helps gain efficiency while still preserving the spatial adaptability that characterizes the \emph{generalized SDPD} model of \cite{DouAlt15}. Moreover, our model allows representation of a wide range of multivariate processes by means of a simple model subject to few assumptions that can be easily checked using real data.

Finally, it is worthwhile to stress the difference between the \emph{SDPD} model of \cite{LeeYu10a}, the \emph{generalized SDPD} model of \cite{DouAlt15}, and the \emph{stationary SDPD} model proposed here. The first model imposes that the spatial relationships be the same for all units, since the coefficients $\lambda_j$ (with $j=0,1,2$) do not change with $i=1,\ldots,p$. Instead, the \emph{stationary SDPD} model in (\ref{b1}) allows varied coefficients for different spatial units, as in the \emph{generalized SDPD} of \cite{DouAlt15}, but they are assumed to be time-invariant thanks to a constraint on the time-lagged parameters. Of course, the estimation procedures vary for the three cases in terms of the convergence rates. The constrained model underlying our \emph{stationary SDPD} allows the estimation procedure to reach the $\sqrt{T}$ convergence rate and to guarantee unbiased estimators, whatever the dimension $p$ and even when $p\rightarrow\infty$ at any rate. This is a big improvement with respect to the other two models. In fact, for the classic \emph{SDPD} model, the estimators are characterized by a $\sqrt{Tp}$ convergence rate (which is faster than that of our model, since they have only three parameters to estimate instead of $2p$), but a bias of order $T^{-1}$ exists, and it does not vanish when $p/T\rightarrow\infty$ (see Theorem 3 of \cite{LeeYu10a}). On the other hand, the convergence rates of the estimators in the \emph{generalized SDPD} model are slower than those of our model and deteriorate when $p\rightarrow\infty$ at a rate faster than $\sqrt{T}$ (see Theorem 2 of \cite{DouAlt15}).

\subsection{Identification of parameters in the case of cross-uncorrelated errors}

In this section, we assume, for simplicity, that the matrix $\bSigma_0^{\varepsilon}=\mathop{var}(\bve_t)$ is diagonal (\emph{i.e.}, there is heteroskedasticity  but no cross-correlation in the error process) and discuss the identifiability of the model. In the next section, we generalize the problem by also adding some cross-correlations in the error process.

Defining $\bz_t^{(0)}=\left[\bI_p-D(\blambda_0)\bW\right]\by_t$, model (\ref{b1}) can be reformulated as
\begin{eqnarray}
\bz_t^{(0)} &=& D(\blambda_1)\bz_{t-1}^{(0)} + \bve_t, \label{b1ter}
\end{eqnarray}
which is a transformed \emph{VAR} process with uncorrelated components, since $D(\blambda_1)$ is diagonal.
Given that we assume that $\bSigma_0^{\varepsilon}$ is also diagonal, the coefficients $\lambda_{1i}$ for $i=1,\ldots,p$, represent the slopes of $p$ univariate autoregressive models with respect to the latent variables $z_{it}^{(0)}$. Therefore, $\lambda_{1i}\equiv\mathop{cor}(z_{it}^{(0)}, z_{i,t-1}^{(0)})$.
From (\ref{b1}), it follows that
\begin{equation}
\left[\bI_p-D(\blambda_0)\bW\right]\bSigma_1 = D(\blambda_1)\left[\bI_p-D(\blambda_0)\bW\right]\bSigma_0 \label{seconda}\\
\end{equation}
and for the $i$-th equation,
\begin{equation}\label{vincolo}
(\be^T_i-\lambda_{0i}\bw^T_i)\bSigma_{1}=\lambda_{1i}(\be^T_i-\lambda_{0i}\bw^T_i)\bSigma_0,
\end{equation}
where $\be_i$ is the column vector with its $i$-th component equal to one and all the others equal to zero, while $\bw_i$ is the column vector containing the $i$-th row of matrix $\bW$.

\begin{figure}
	\centering
	\mbox{ \resizebox{4.8cm}{!}{\includegraphics{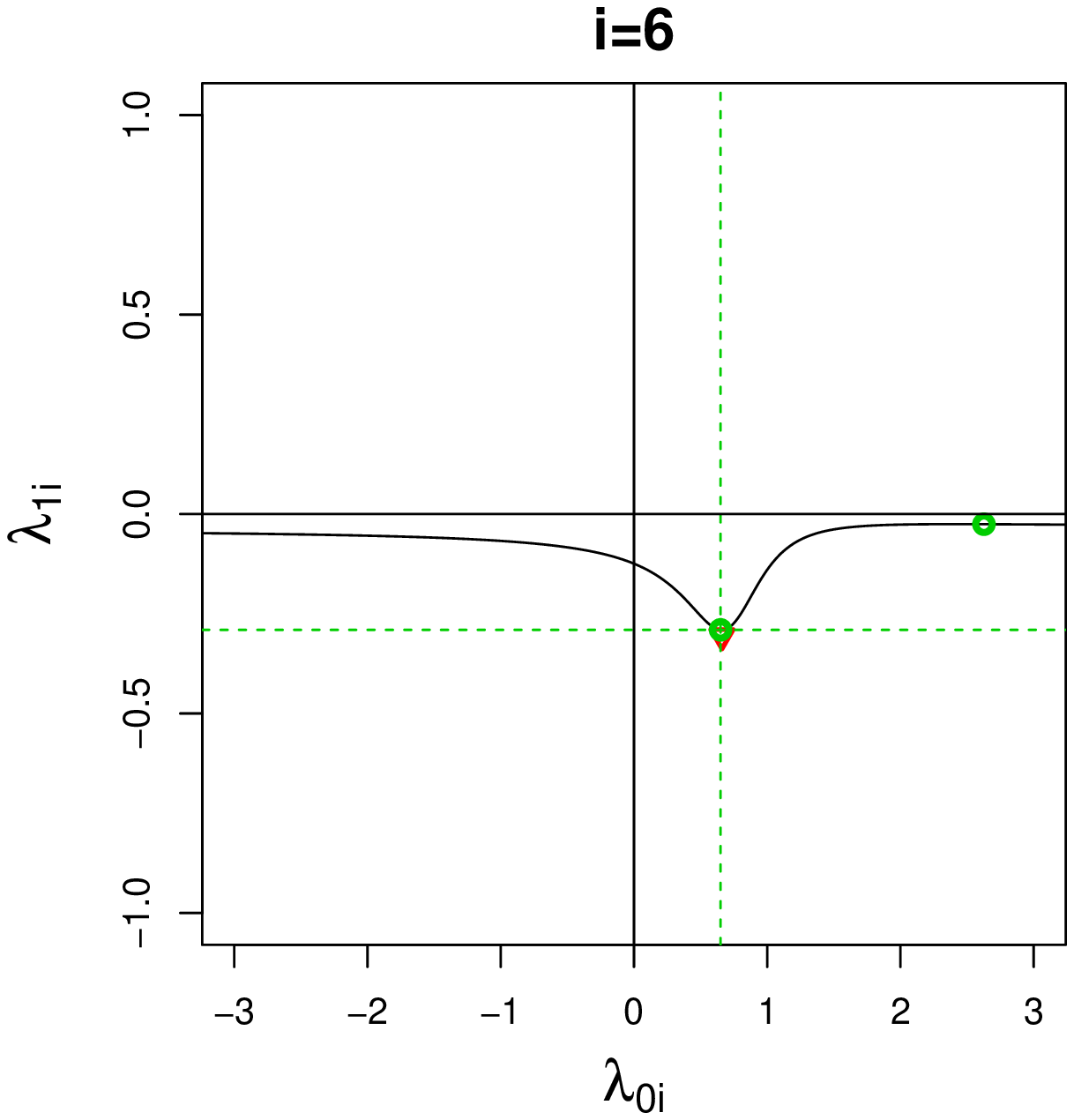}} }
	\mbox{ \resizebox{4.8cm}{!}{\includegraphics{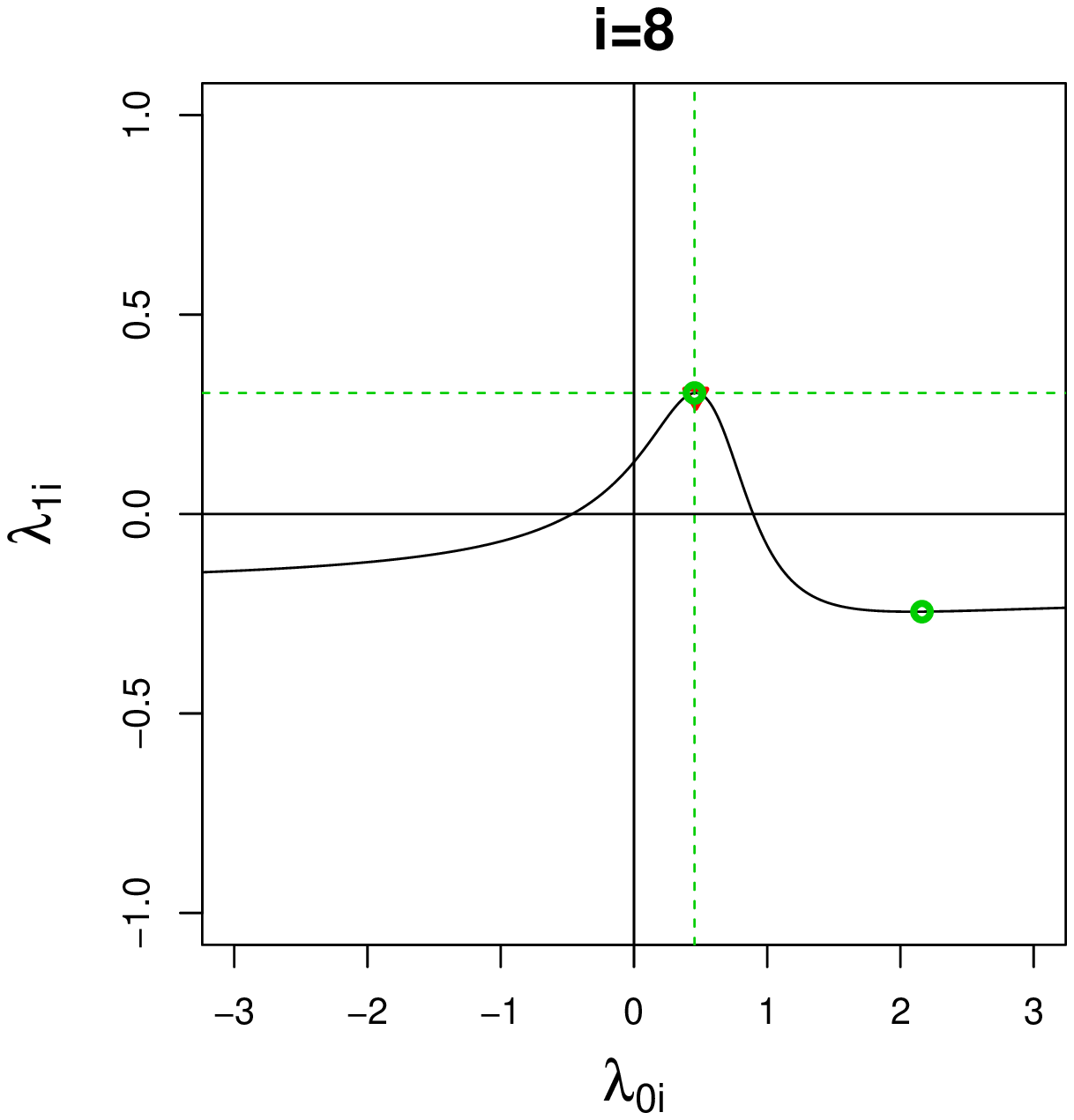}} }
	\mbox{ \resizebox{4.8cm}{!}{\includegraphics{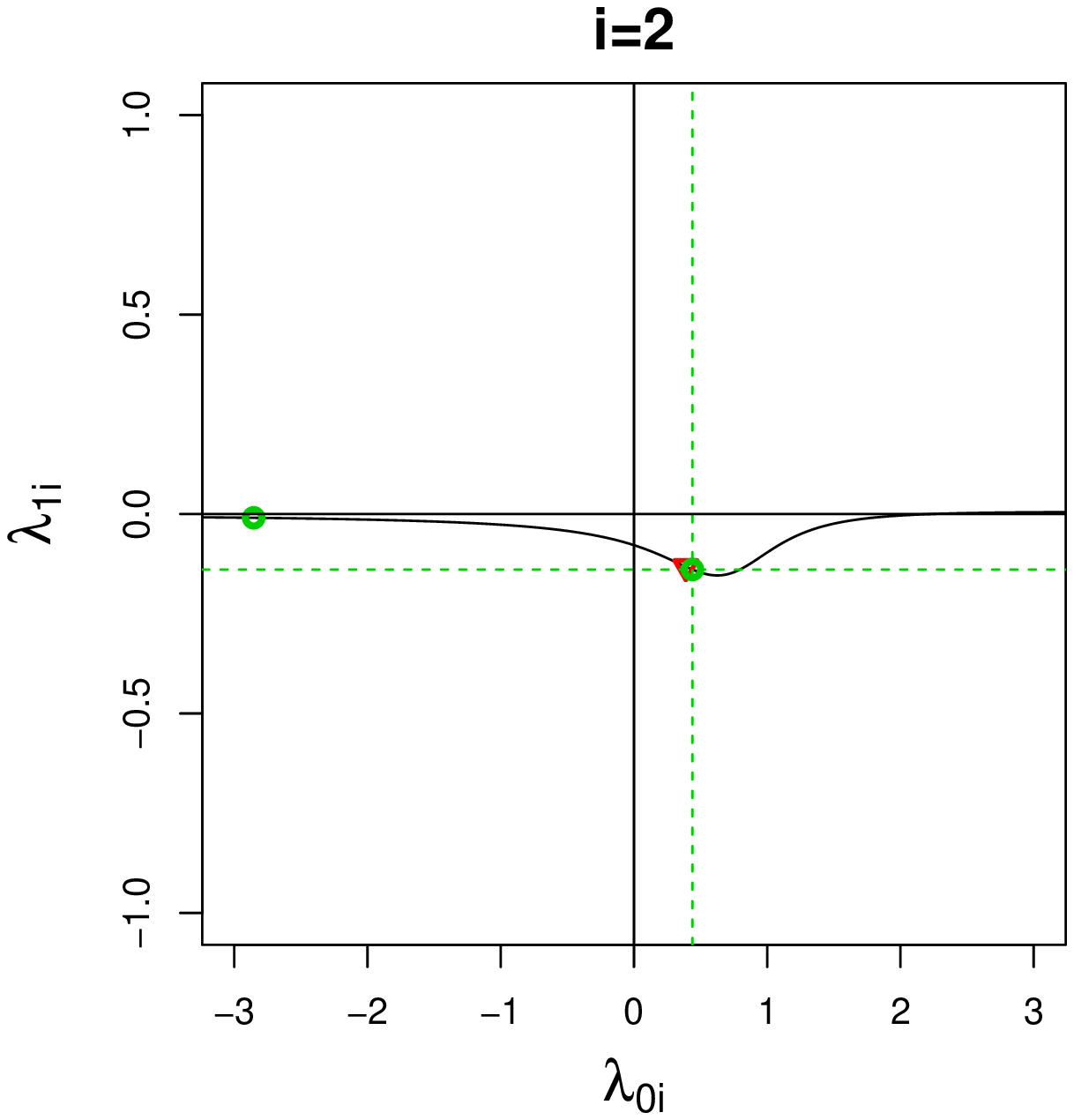}} }
\caption{\label{figure2} \small Plots of $\lambda_{1i}=\mathop{cor}(z_{it}^0, z_{i,t-1}^0)$ as a function of $\lambda_{0i}$ for three different locations ($i=2,6,8$) of model 1 used in section \ref{simulazioni}. The dots show the two candidate solutions for the estimation. Among these, the vertical and horizontal dashed lines identify the one that satisfies the sufficient condition in (\ref{vincolo}). As expected, it coincides with the true values $(\lambda^*_{0i},\lambda^*_{1i})$ of model 1. Cases $i=6,8$ refer to locations with cross-correlated errors while case $i=2$ represents a location with cross-uncorrelated errors.}
\end{figure}

Under general assumptions, (\ref{vincolo}) admits only one solution with respect to $\lambda_{0i}$ and $\lambda_{1i}$ (see Theorem \ref{theorem1}), which can be found among the extreme points of $\lambda_{1i}=\mathop{cor}(z_{it}^{(0)}, z_{i,t-1}^{(0)})$ as a function of $\lambda_{0i}$.
To provide insight into this, the first two plots of figure \ref{figure2} show two examples based on model 1 used in the simulation study. Denote with ($\lambda_{0i}^*,\lambda_{1i}^*)$ the true values of the coefficients used in model 1 for a given location $i$ (in particular, in figure \ref{figure2}, the first two plots refer to locations $i=6$ and $i=8$). The solid line shows $\lambda_{1i}=\mathop{cor}(z_{it}^{(0)}, z_{i,t-1}^{(0)})$ as a function of $\lambda_{0i}$. The two dots show the points of this function where the first derivative is zero. The vertical and horizontal dashed lines identify which one of the two points satisfies the sufficient condition in (\ref{vincolo}). As expected, it coincides with the true values ($\lambda_{0i}^*,\lambda_{1i}^*)$ used to generate the time series.
Theorem \ref{theorem1}, shown in the Appendix, formalizes this result.
\begin{theorem}\label{theorem1}
Consider model (\ref{b1}) for a stationary process $\by_t$ with mean zero, and assume that the error process $\bve_t$ is such that $\bSigma^0_\varepsilon=\mathop{var}(\bve_t)$ is diagonal (\emph{i.e.}, there is heteroskedasticity but no cross-correlation in the errors). Under assumptions $A1-A4$ in section \ref{asymptotic}, the following results hold:
\begin{enumerate}
\item There exist a unique couple of values $(\lambda_{0i}^*,\lambda_{1i}^*)$ satisfying the following system of equations:
\begin{equation}\label{vinc}
(\be^T_i-\lambda_{0i}\bw^T_i)\bSigma_{1}-\lambda_{1i}(\be^T_i-\lambda_{0i}\bw^T_i)\bSigma_0={\bf 0}^T, \qquad\qquad i=1,\ldots,p,
\end{equation}
where $\be_i$ is the $i$-th unit vector, and $\bw_i$ contains the $i$-th row of the spatial matrix $\bW$.
\item Such a point, $(\lambda_{0i}^*,\lambda_{1i}^*)$, is also the solution of the following second-order polynomial equation:
\begin{eqnarray} \label{nec_cond}
\left.\frac{\partial\mathop{cov}(z_{it}^{(0)}, z_{i,t-1}^{(0)})}{\partial\lambda_{0i}}\right|_{\lambda_{0i}=\lambda_{0i}^*}- \lambda_{1i}^*\left.\frac{\partial\mathop{var}(z_{i,t-1}^{(0)})}{\partial\lambda_{0i}}\right|_{\lambda_{0i}=\lambda_{0i}^*} &=& 0.
\end{eqnarray}
\end{enumerate}
\end{theorem}

\noindent\textbf{Remark 1:} Theorem \ref{theorem1} not only shows that the \emph{stationary SDPD} model is well identified, because there is a unique solution for $(\blambda_0,\blambda_1)$, but it also suggests a way to estimate such parameters. In fact, we can find all the solutions to equation (\ref{nec_cond}) and then check which one satisfies the sufficient condition in (\ref{vincolo}). This estimation procedure is described in section \ref{est_alg}.

\subsection{Identification of parameters in the case of cross-correlated errors}

Now, we relax the assumption on the error $\bve_t$ by letting $\bSigma_0^\varepsilon$ be a full matrix (i.e., there is heteroskedasticity and cross-correlation in the error process). This setup allows the process $\by_t$ to include some {spurious cross-correlation} not explained by $\bW$. In this case, the coefficients $\lambda_{i1}$ still give the correlations between the latent variables $z_{i,t}^{(0)}$ and $z_{i,t-1}^{(0)}$, but now, the $p$ equations in model (\ref{b1ter}) are correlated. The main consequence of this is that the true values $(\lambda_{0i}^*, \lambda_{1i}^*)$ do not identify an extreme point of the correlation function (see case $i=2$ in figure \ref{figure2}). Anyway, the sufficient condition in (\ref{vincolo}) is still valid, and the true coefficients $(\lambda_{0i}^*, \lambda_{1i}^*)$ can be identified by introducing a ``constrained'' condition.

\begin{theorem}\label{theorem1bis}
Consider model (\ref{b1}) for a stationary process $\by_t$ with mean zero, and assume that the error process $\bve_t$ is such that $\bSigma^0_\varepsilon=\mathop{var}(\bve_t)$ is a full matrix (\emph{i.e.}, there is heteroskedasticity and cross-correlation in the errors). Under assumptions $A1-A4$ in section \ref{asymptotic}, the following results hold:
\begin{enumerate}
\item There exist a unique couple of values $(\lambda_{0i}^*,\lambda_{1i}^*)$ satisfying the following system of equations
\[
(\be^T_i-\lambda_{0i}\bw^T_i)\bSigma_{1}-\lambda_{1i}(\be^T_i-\lambda_{0i}\bw^T_i)\bSigma_0={\bf 0}^T, \qquad\qquad i=1,\ldots,p,
\]
where $\be_i$ is the $i$-th unit vector, and $\bw_i$ contains the $i$-th row of the spatial matrix $\bW$;
\item such a point, $(\lambda_{0i}^*,\lambda_{1i}^*)$, is also the solution of the following second-order polynomial equation.
\begin{eqnarray*}
\left.\frac{\partial\mathop{cov}(z_{it}^{(0)}, z_{i,t-1}^{(0)})}{\partial\lambda_{0i}}\right|_{\lambda_{0i}=\lambda_{0i}^*}- \lambda_{1i}^*\left.\frac{\partial\mathop{var}(z_{i,t-1}^{(0)})}{\partial\lambda_{0i}}\right|_{\lambda_{0i}=\lambda_{0i}^*} &=& (\be_i^T-\lambda_{0i}^*\bw^T_i)(\bSigma_1^T-\bSigma_1)\bw_i.
\end{eqnarray*}
\end{enumerate}
\end{theorem}

\noindent\textbf{Remark 2:} When the errors $\bve_t$ are not cross-correlated, the matrix $\bSigma_1$ is symmetric by Lemma \ref{lemma1} in the Appendix, so that point 2 in Theorem \ref{theorem1bis} becomes the same as in Theorem \ref{theorem1}. Therefore, Theorem \ref{theorem1bis} includes Theorem \ref{theorem1} as a special case.

\section{Estimation procedure}\label{est_alg}
We present here a simple algorithm for the estimation of the parameters $(\lambda_{0i},\lambda_{1i})$ for $i=1,\ldots,p$. First, estimate the matrices $\bSigma_1$ and $\bSigma_0$ through some consistent estimators $\hat\bSigma_0$ and $\hat\bSigma_1$. For example, $\hat\bSigma_0=(n-1)^{-1}\bY_0\bY_0^T$ and $\hat\bSigma_1=(n-2)^{-1}\bY_0\bY_1^T$, where $\bY_l=(\by_{1+l}, \cdots, \by_{n-l})$. Alternatively, the threshold estimator analyzed in \cite{CheAlt13} can be considered in the high dimensional setup. Then, for each location $i=1,\ldots,p$, implement the following steps.
\begin{enumerate}
\item Define $\be_i$ as the $i$-th unit vector and $\bw^T_i=\be^T_i\bW$, then compute:
\begin{eqnarray*}
\hat a_{0i} &=& \be_i^T\hat\bSigma_0\be_i, \quad \hat a_{1i} = \be_i^T\hat\bSigma_1 \be_i, \quad
\hat a_{2i} = \be_i^T(\hat\bSigma_1^T-\hat\bSigma_1)\bw_i, \\
\hat b_{0i} &=& -2\be_i^T\hat\bSigma_0\bw_i, \quad \hat b_{1i} =  -\be^T_i(\hat\bSigma_1+\hat\bSigma_1^T)\bw_i, \\
 \hat c_{0i} &=& \bw^T_i\hat\bSigma_0\bw_i,\quad \hat c_{1i} = \bw^T_i\hat\bSigma_1\bw_i.
\end{eqnarray*}
\item Find the two roots $\lambda_{0i}^{(1)}$ and $\lambda_{0i}^{(2)}$ of the following two-order polynomial equation.
\begin{equation}\label{eqq}
\hat t_{0i} + \hat t_{1i}\lambda_{i0} + \hat t_{2i}\lambda_{i0}^2 =0,
\end{equation}
where $\hat t_{0i} = \hat b_{1i}\hat a_{0i}-\hat b_{0i}\hat a_{1i}+\hat a_{0i}\hat a_{2i}$, $\hat t_{1i} = 2(\hat a_{0i}\hat c_{1i}-\hat c_{0i}\hat a_{1i})+\hat a_{2i}\hat b_{0i}$, and $\hat t_{2i} = \hat c_{1i}\hat b_{0i}-\hat c_{0i}\hat b_{1i}+\hat a_{2i}\hat c_{0i}$.
\item Estimate $\lambda_{0i}$ and $\lambda_{1i}$ by
\begin{eqnarray}\label{stimatore0}
\hat\lambda_{0i}&=&\arg\min_{j=1,2}\bv_{ij}^T\bv_{ij}, \\
\hat\lambda_{1i}&=&\frac{(\be_i^T-\hat\lambda_{0i}\bw^T_i)\hat\bSigma_1(\be_i-\hat\lambda_{0i}\bw_i)}{(\be_i^T-\hat\lambda_{0i}\bw^T_i)\hat\bSigma_0(\be_i-\hat\lambda_{0i}\bw_i)}, \label{stimatore1}
\end{eqnarray}
where
$
\bv^T_{ij} = (\be_i^T-\lambda_{0i}^{(j)}\bw^T_i)\hat\bSigma_1-\lambda_{1i}^{(j)}(\be_i^T-\lambda_{0i}^{(j)}\bw^T_i)\hat\bSigma_0,
$,
and $\lambda_{1i}^{(j)} = (\be_i^T-\lambda_{0i}^{(j)}\bw^T_i)\hat\bSigma_1(\be_i-\lambda_{0i}^{(j)}\bw_i)/(\be_i^T-\lambda_{0i}^{(j)}\bw^T_i)\hat\bSigma_0(\be_i-\lambda_{0i}^{(j)}\bw_i)$.
\end{enumerate}


\vspace{10pt}\noindent\textbf{Remark 3:} Assumption $A2$ in section \ref{asymptotic} guarantees that matrix $\left[\bI_p-D(\blambda_0)\bW\right]$ has full rank. However, the above estimation procedure may suffer for some locations if matrix $\left[\bI_p-D(\blambda_0)\bW\right]$ is near singularity. Such a case may come about because of the presence of some almost linearly dependent rows in the matrix, which may cause the quantity $\bw_i^T[\bSigma_1-\lambda_{1i}\bSigma_0]\bw_i$ to be almost zero for those rows (see Lemma \ref{lemma2}). As a result, the procedure loose efficiency for the estimation of $\lambda_{i0}$ for those locations (but it still works for $\lambda_{1i}$). Something similar may happen if there are some (almost) zero rows in $\bW$, which is excluded by assumption $A4$. Anyway, it is worthwhile to stress that the estimation procedure works efficiently for all the other locations. In fact, the procedure does not require the inversion of matrix $\left[\bI_p-D(\blambda_0)\bW\right]$, so it is able to isolate and separate the effects of ``collinear'' locations (or uncorrelated locations) from the other locations and to guarantee consistent and efficient estimations for the last locations.

\section{Theoretical results}\label{asymptotic}

In this section, we show the theoretical foundations of our proposal. In particular, we present the assumptions and show the consistency and the asymptotic normality of the estimators, for the cases of finite dimension and high dimension. Moreover, we show that the \emph{stationary SDPD} model can be used to represent a wide range of multivariate linear processes with respect to a ``latent spatial matrix,'' and therefore, it is of wider application than classic spatio--temporal contexts.

The reduced form of model (\ref{b1}) is
\begin{equation}\label{b1bis}
\by_t=\bA^*\by_{t-1}+\bve_t^*,
\end{equation}
where $\bve_t^*=\left[\bI_p-D(\blambda_0)\bW\right]^{-1}\bve_t$ and
\begin{equation}\label{diagonalize}
\bA^*=\left[\bI_p-D(\blambda_0)\bW\right]^{-1}D(\blambda_1)\left[\bI_p-D(\blambda_0)\bW\right].
\end{equation}
Note that the errors $\bve_t^*$ have mean zero and are serially uncorrelated. Model (\ref{b1bis}) has a \emph{VAR} representation, so it is stationary when all the eigenvalues of matrix $\bA^*$ are smaller than one in absolute value. From (\ref{diagonalize}), we can note that $\blambda_1$ contains the eigenvalues of $\bA^*$ whereas $\blambda_0$ only affects its eigenvectors (see the proof of Theorem \ref{theorem1}). Therefore, we must consider the following assumptions:
\begin{itemize}
\item[A1)] $\lambda_{1i}\in\mathbb{R}$ and $|\lambda_{1i}|<1$, for all $i$, and vector $\blambda_1$ is not scalar;
\item[A2)] $\lambda_{0i}\in\mathbb{R}$ for all $i$ and vector $\blambda_0$ is such that matrix $\left[\bI_p-D(\blambda_0)\bW\right]$ has full rank;
\item[A3)] the errors $\varepsilon_{it}$ are serially independent and such that $E(\varepsilon_{it})=0$ and $E|\varepsilon_{it}|^\delta<\infty$ for all $i,t$, for some $\delta>4$;
\item[A4)] the spatial matrix $\bW$ is nonsingular and has zero main diagonal.
\end{itemize}

Assumption $A1$ implies stationarity. Moreover, it guarantees that there are at least two distinct values in vector $\blambda_1$ so that model (\ref{b1}) is identifiable, as shown in Theorem \ref{theorem1}. Assumption $A2$ is clearly necessary to assure that matrix $\left[\bI_p-D(\blambda_0)\bW\right]$ can be inverted so that the reduced model in (\ref{b1bis}) is well defined (Remark 3 indicates what happens when this assumption is not satisfied). Incidentally, it is worthwhile to note that our setup automatically solves the problem concerning the parameter space of $\blambda_0$, highlighted at the end of section 2.2 by \cite{KelPru10}. So, in the empirical applications of our model, it is possible to use any kind of normalization for $\bW$ (\emph{i.e.}, row-factor normalization or single-factor normalization), since the vector $\blambda_0$ would automatically rescale accordingly (see section \ref{high} for more details on this aspect). This means that the coefficients $\lambda_{0i}$ can also take values outside the classic interval $[-1,1]$.
Assumption $A3$ assures that the results in \cite{Han76} can be applied to show the asymptotic normality of the estimators.
Assumption $A4$ is classic in spatio--temporal models and guarantees that the model is well defined and identifiable with respect to all the parameters, also for $p\rightarrow\infty$ (see Lemma \ref{lemma2} and Theorem \ref{theorem4}).

Under assumptions $A1-A4$, it is immediately evident that the estimators $\hat\blambda_{0}$ and $\hat\blambda_{1}$, presented in section \ref{est_alg}, are both consistent following Theorem 11.2.1 in \cite{BroDav86}. For asymptotic normality, the following theorem can be stated.
\begin{theorem}\label{theorem3}
Consider $\hat\lambda_{0i}$ and $\hat\lambda_{1i}$, the estimators obtained by the algorithm in section \ref{est_alg}. Under assumptions $A1-A4$, we have for finite $p$
\begin{equation}
\sqrt{T}(\hat\lambda_{ji}-\lambda_{ji}) \stackrel{d}{\longrightarrow}N(0,\bD_{ji}^T\bV_{ji}\bD_{ji}) \nonumber\qquad\qquad  j=0,1;\quad i=1,\ldots,p,
\end{equation}
where $\bD_{ji}$ are the $K_i\times 1$ vectors, and $\bV_{ji}$ are the matrices of order $K_i$ with $K_i \le 2p^2$ (see the proof).
\end{theorem}

Note that the estimators $\widehat\lambda_{ji}$ are unbiased for all $i,j$ and for all $p$. 

In the high dimension, we have infinite parameters to estimate ($2p$ in total, where $p\rightarrow\infty$). Therefore, we must assure that the consistency of the estimators is still valid in such a case. As expected, the properties of matrix $\bW$ influence the consistency and the convergence rates of the estimators $\hat\lambda_{ij}$ when $p\rightarrow\infty$. For example, denote with $k_i$ the number of nonzero elements in vector $\bw_i$. If $k_i=O(1)$ as $p\rightarrow\infty$, for all $i$, then the effective dimension of model (\ref{b1}) is finite and Theorem \ref{theorem3} can still be applied for the consistency and the asymptotic normality of the estimators $\hat\lambda_{ji}$, even if $p\rightarrow\infty$. The following Theorem \ref{theorem4}, instead, shows the consistency of the estimators under more general vectors $\bw_i$, with $k_i\rightarrow\infty$ as $p\rightarrow\infty$. 

\subsection{Asymptotics for high dimensional setups}\label{high}
In model (\ref{b1}), the spatial correlation between a given location $i$-th and the other locations is summarized by $\lambda_{0i}\bw_i$. If the vector $\bw_i$ is rescaled by a factor $\delta_i$, then we can have an equivalent model by rescaling the spatial coefficient $\lambda_{0i}$ by the inverse of the same factor, since $\lambda_{0i}\bw_i=\delta_i^{-1}\lambda_{0i}\bw_i\delta_i=\lambda_{0i,\delta}\bw_{i,\delta}$. In such a way, we may consider irrelevant a row-normalization of matrix $\bW$ if we let the coefficients in  $D(\blambda_{0})$ rescale accordingly. Such an approach is not new and follows the idea of \cite{KelPru10}. We use this approach here in order to simplify the analysis and the interpretation of the \emph{stationary SDPD} model in the high dimensional setup.

In fact, when $p\rightarrow\infty$ and $k_i=O(p)$, the vectors $\bw_i$ may change with $p$ and this may have an influence on the scale order of the process.
This happens, for example, if we consider a row-normalized spatial matrix $\bW$, since the weights become infinitely small for infinitely large $p$. Looking at the (\ref{diagonalize}), model (\ref{b1}) appears to become spatially uncorrelated for $p\rightarrow\infty$ because matrix $\bW$ tends to be asymptotically diagonal (for $p\rightarrow\infty$ and $T$ given). As a consequence, the model appears to become not identifiable in the high dimension with respect to the parameters $\lambda_{0i}$. To avoid this, we assume here that also the coefficients $\lambda_{0i}$ may depend on the dimension $p$, borrowing the idea of \cite{KelPru10}. In such a way, we can derive the conditions for the identifiability of the model in the high dimension and better convergence rates for the estimators. This is shown by the following theorem.

\begin{theorem}\label{theorem4}
Consider $\hat\lambda_{0i}$ and $\hat\lambda_{1i}$, the estimators obtained by the algorithm in section \ref{est_alg}. Assume that the number of nonzero values in $\bw_i$ is $k_i=O(p)$ for all $i=1,\ldots,p$. Under assumptions $A1-A4$, for $p\rightarrow\infty$ we have the following cases:
\begin{itemize}
\item[(i)] if the vectors $\bw_i$ are normalized by $L_1$ norm then
\[
\left|\hat\lambda_{ji}-\lambda_{ji}\right| =O_p(T^{-1/2}) \nonumber\qquad\qquad  {\rm for\ } j=0,1;i=1,\ldots,p,
\]
provided that $\lambda_{0i}=O(p)$;
\item[(ii)] if the vectors $\bw_i$ are normalized by $L_2$ norm and $\lambda_{0i}=O(1)$ then
\[
\left|\hat\lambda_{ji}-\lambda_{ji}\right| =O_p(T^{-1/2}) \nonumber\qquad\qquad  {\rm for\ } j=0,1;i=1,\ldots,p;
\]
\item[(iii)] for generic (not normalized but bounded) vectors $\bw_i$ and $\lambda_{0i}=O(1)$ we have
\[
\left|\hat\lambda_{ji}-\lambda_{ji}\right| =O_p(pT^{-1/2}) \nonumber\qquad\qquad  {\rm for\ } j=0,1;i=1,\ldots,p.
\]
\end{itemize}
\end{theorem}

As shown by Theorem \ref{theorem4}, cases (i) and (ii), if we consider a row-normalized spatial matrix $\bW$, our estimation procedure is consistent for any value of $p$ and with $p\rightarrow\infty$ at any rate. In other words, the convergence rate is not affected by the dimension $p$. However, there are some differences between the two cases of $L_1$ and $L_2$ normalization. In the first case, we need to impose that the spatial coefficients $\lambda_{0i}$ increases in the order $O(p)$ as $p\rightarrow\infty$ (otherwise the model becomes not identifiable in the high dimension), whereas in the last case of $L_2$ norm they can remain constant for $p\rightarrow\infty$.
In case (iii), which is more general because it is valid for any $\bW$, we need to impose $k_i=o(T^{1/2})$ in order to guarantee the consistency of the estimators.

To complete this section, we want to show the class of processes that can be analysed by our \emph{stationary SDPD} model. Under assumption $A2$, any \emph{stationary SDPD} model can be equivalently represented as a VAR process as in (\ref{b1bis}), with respect to an autoregressive matrix coefficient $\bA^*$ defined in (\ref{diagonalize}).
Now, by exploiting the simple structure of our model, we can show the conditions under which the opposite is true. The following corollary derives from standard results.
\begin{corollary}\label{corollary1}
Given a stationary multivariate process $\by_t=\bA^*\by_{t-1}+\bve_t^*$, with $\bve_t^*$ satisfying assumption $A3$, a necessary and sufficient condition to represent the process $\by_t$ by a stationary SDPD model is that matrix $\bA^*$ is diagonalizable. Therefore, matrix $\bA^*$ must have $p$ linearly independent eigenvectors. This is (alternatively) assured by one of the following sufficient conditions:
\begin{itemize}
\item the eigenvalues $\lambda_{11},\ldots,\lambda_{1p}$ of matrix $\bA^*$ are all distinct, or
\item the eigenvalues $\lambda_{11},\ldots,\lambda_{1p}$ of matrix $\bA^*$ consist of $h$ distinct values $\mu_1,\ldots,\mu_h$ having geometric multiplicities $r_1,\ldots,r_h$, such that $r_1+\ldots+r_h=p$.
\end{itemize}
\end{corollary}
By corollary \ref{corollary1} and assumptions $A1-A4$, the VAR processes that cannot be represented and consistently estimated by our \emph{stationary SDPD}  model are those characterized by a matrix $\bA^*$ with linear dependent eigenvectors (i.e., those with algebraic multiplicities) or those with complex eigenvalues. In order to apply our model to those cases also, we should generalize the estimation procedure using the Jordan decomposition of matrix $\bA^*$. However, we leave this topic to future study.

\section{Simulation study}\label{simulazioni}

This section contains the results of a simulation study implemented to evaluate the performance of the proposed estimation procedure. In section 5.1, we describe the settings and check the validity of the assumptions for the simulated models. Then, in section 5.2, we evaluate the consistency of the estimation procedure and the convergence rate for the estimators using a known spatial matrix. Finally, section 5.3, we analyse the case when the spatial matrix $\bW$ is unknown, and therefore, to be estimated.

\subsection{Settings}

We consider three different spatial matrices. In the first, we randomly generate a matrix of order $p\times p$, and we post-multiply this matrix by its transpose in order to force symmetry. The resulting spatial matrix is denoted with $\bW_1$. Note that such a matrix is \emph{full}, and it may have positive and negative elements. In the other two cases, the spatial matrix is \emph{sparse} and has only positive entries: $\bW_2$ is generated by setting to one only four values in each row while $\bW_3$ is generated by setting to one $2\sqrt{p}$ elements in each row.
For all three cases, we check the rank to guarantee that the spatial matrix has $p$ linearly independent rows. Moreover, we set to zero the main diagonal, and we rescale the elements so that each row has norm equal to one ($L_2$ row-normalization).

For the error process, we generate $p$ independent Gaussian series $e_{ti}$ with mean zero and standard error ${\sigma}_i$, where the values ${\sigma}_i$ are generated randomly from a uniform distribution $U(0.5, 1.5)$ for $i=1,\ldots,p$. Then, we define the cross-correlated error process $\bve_t=\{\varepsilon_{it},t=1,\ldots,T\}$, where
\[
\left\{
\begin{array}{ll}
\varepsilon_{ti} = e_{ti} -0.7*e_{t2} & \mbox{for }i=3,\ldots,p, \\
\varepsilon_{ti} = e_{ti} & \mbox{otherwise}. \\
\end{array} \right.
\]

We generate all $\lambda_{ji}$ from a uniform distribution $U(-0.7, 0.7)$. The settings above guarantee that assumptions $A1-A4$ hold. We generate different models with dimensions $p = (10, 50, 100, 500)$ and sample sizes $T = (50, 100, 500, 1000)$. Note that we may have $T<<p$. For each configuration of settings, we generate 500 Monte Carlo replications of the model and report the estimation results. All the analyses have been made in R.


\subsection{Empirical performance of the estimators when $\bW$ is known}

\begin{figure}
	\centering
	\mbox{ \resizebox{15cm}{!}{\includegraphics{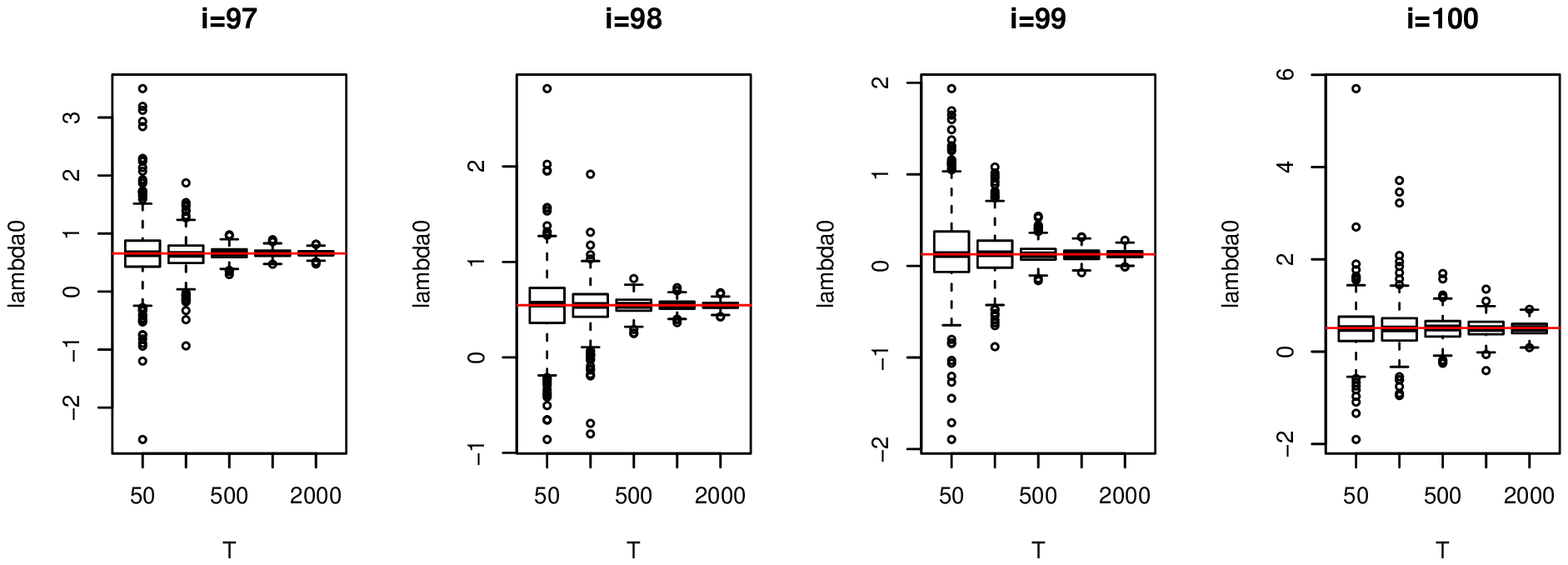}} }
	\mbox{ \resizebox{15cm}{!}{\includegraphics{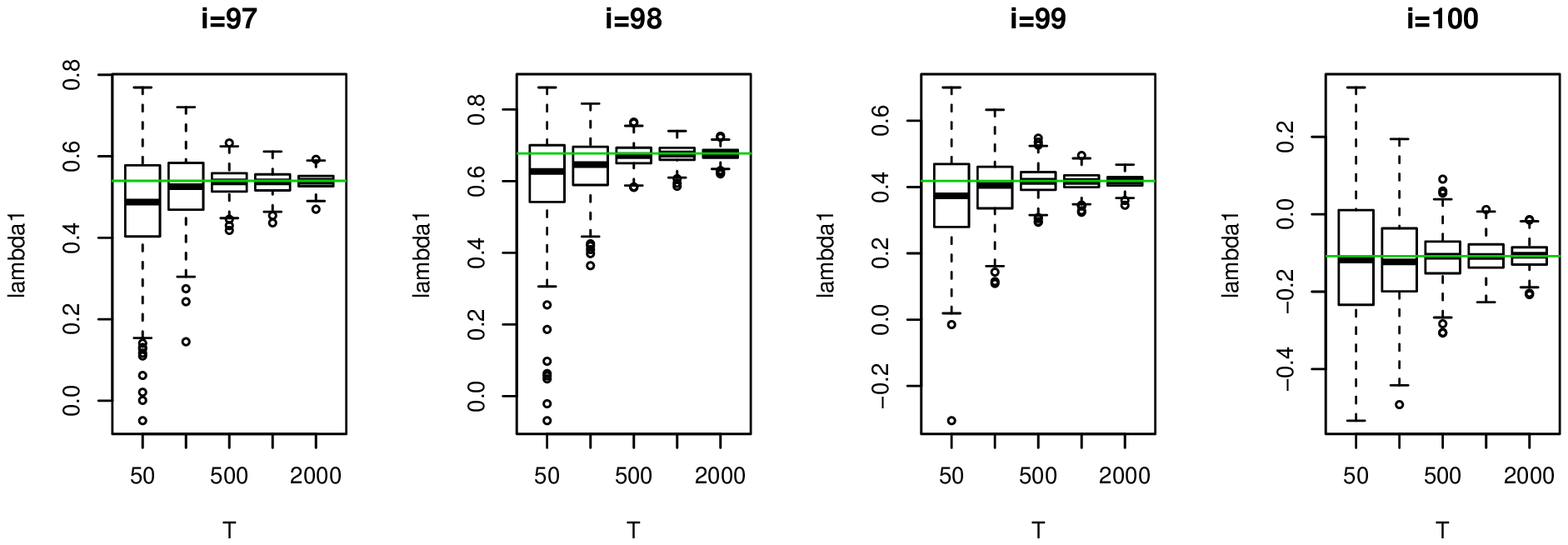}} }
\caption{\label{figure5} \small Estimations of the coefficients $\lambda_{0i}$ (at the top) and $\lambda_{1i}$ (at the bottom) for different time series length $T$ and fixed dimension $p=100$. Each plot focuses on a different {location} $i$ with $i=97,\ldots,100$. The true values of the coefficients $\lambda_{ji}$ are shown through the horizontal solid lines.}
\end{figure}

Figure \ref{figure5} shows the box plots of the estimations for increasing sample sizes $T = (50, 100, 500, 1000)$ and fixed dimension $p = 100$. The four plots at the top refer to the estimation of $\lambda_{0i}$ while the four plots at the bottom refer to that of $\lambda_{1i}$. Each plot focuses on a different {location} $i$, where $i = 97,\ldots,100$. The true values of the coefficients $\lambda_{ji}$ are shown through the horizontal lines. Note that we have $T\leq p$ for the first two box plots in each plot, since $p = 100$ for this model. The box plots are centred on the true value of the parameters, and the variance reduces for increasing values of $T$, showing consistency of the estimators and a good performance for small $T$/large $p$ also.

To evaluate the estimation error, for each realized time series, we compute the average error $AE$ and the average squared error $ASE$ using the equations below.
\begin{equation}\label{mse}
AE(\widehat\blambda_j) = \frac{1}{p}\sum_{i=1}^p{(\hat\lambda_{ji}-\lambda_{ji})}, \qquad ASE(\widehat\blambda_j) = \frac{1}{p}\sum_{i=1}^p{(\hat\lambda_{ji}-\lambda_{ji})^2}, \qquad j=1,2.
\end{equation}
Table \ref{tabella1} reports the mean values of $ASE(\widehat\blambda_j)$ (with the standard deviations in brackets) computed over 500 simulated time series for different values of $T$ and $p$.

\begin{figure}
	\centering
	\mbox{ \resizebox{15cm}{!}{\includegraphics{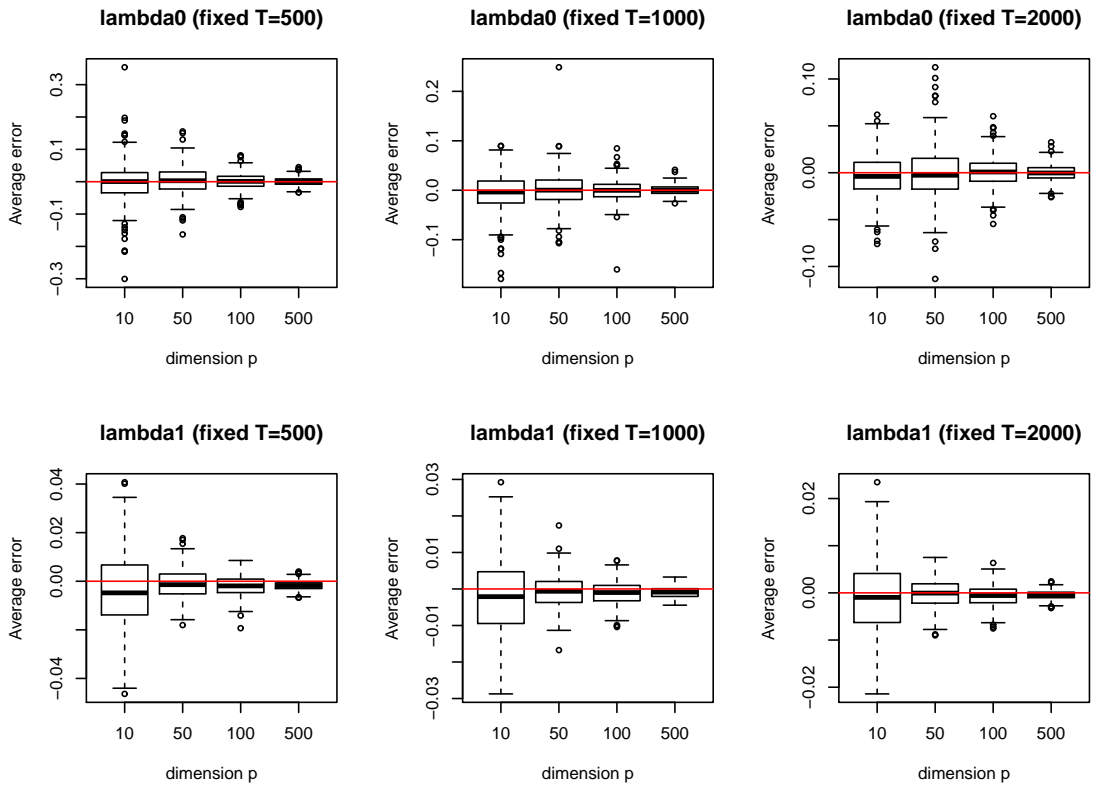}} }
\caption{\label{increasing_p_global} \small Distribution of the average errors $AE(\blambda_0)$ (at the top) and $AE(\blambda_1)$ (at the bottom) from 500 replications of the model based on the spatial matrix $\bW_1$, with different values of $p$ and sample sizes $T$ (each plot is for a given $T$). }
\end{figure}

As shown in the table, the estimation error decreases when the sample size $T$ increases. It is interesting to note that the estimation error does not increase for increasing values of the dimension $p$. This is more evident from figure \ref{increasing_p_global}, which shows the box plots of the average errors $AE(\blambda_0)$ (at the top) and $AE(\blambda_1)$ (at the bottom) computed for 500 replications of the model, with varying values of $p$, sample sizes $T$, and spatial matrix $\bW_1$. We can note from the figure that $\hat\blambda_0$ and $\hat\blambda_1$ are unbiased for all $n$ and $p$. Moreover, the variability of the box plots decreases for $p\rightarrow\infty$ and fixed $T$: this is a consequence of averaging the absolute error over the $p$ locations using equation (\ref{mse}).


\subsection{Estimation results when the spatial matrix is unknown}\label{matrixA}
In this section, we evaluate the performance of the proposed estimation procedure when the spatial matrix $\bW$ is unknown and needs to be estimated.
In this case, the estimation error has to be evaluated with respect to matrix $\bA^*$ in order to include the effects of both $\hat\blambda_{j}$ and $\hat\bW$ on the final estimations. So, using (\ref{diagonalize}), we define the two estimators
\begin{eqnarray}
\hat\bA_{SDPD}^*(\bW) &=& \left[\bI_p-D(\hat\blambda_0)\bW\right]^{-1}D(\hat\blambda_1)\left[\bI_p-D(\hat\blambda_0)\bW\right] and \label{AW}\\
\hat\bA_{SDPD}^*(\hat\bW) &=& \left[\bI_p-D(\hat\blambda_0)\hat\bW\right]^{-1}D(\hat\blambda_1)\left[\bI_p-D(\hat\blambda_0)\hat\bW\right],\label{AWhat}
\end{eqnarray}
where matrix $\bW$ is assumed to be known in the first case and unknown in the second. When $\bW$ is unknown, we estimate it by the (row-normalized) correlation matrix at lag zero, but other more efficient estimators of $\bW$ can be considered alternatively.

For the sake of comparison, remembering the \emph{VAR} representation of our model in (\ref{b1bis}), we also estimate matrix $\bA^*$ using the classic Yule--Walker estimator of the VAR model $\hat\bA_{VAR}^*=\hat\bSigma_0^{-1}\hat\bSigma_1$.

\begin{figure}
	\centering
	\mbox{ \resizebox{14cm}{!}{\includegraphics{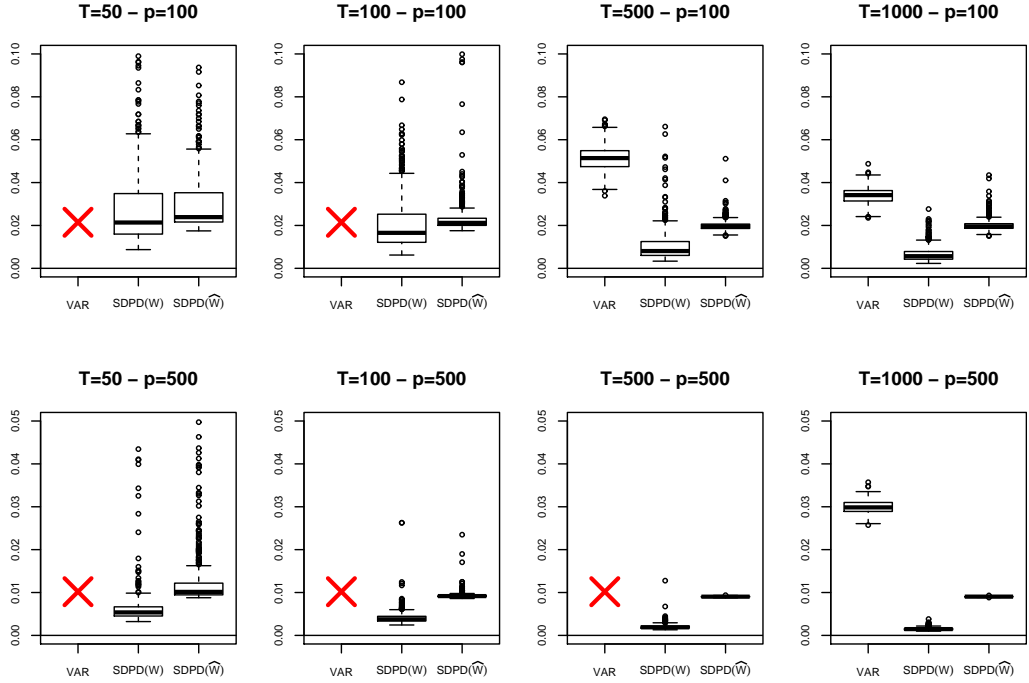}} }
\caption{\label{figure6} \small average squared error for the estimation of $\bA^*$ computed by (\ref{mse2}) for different values of $p$ and $T$. The three box plots in each plot compare three different estimators: the classic Yule--Walker estimator of the VAR model on the left, our estimator $\hat\bA_{SDPD}^*(\bW)$ defined in (\ref{AW}) for an SDPD model with the known spatial matrix in the middle, and our estimator $\hat\bA_{SDPD}^*(\hat\bW)$ defined in (\ref{AWhat}) for an SDPD model with the estimated spatial matrix on the right. Note that the classic Yule--Walker estimator of the VAR model cannot be computed when $T\leq p$.}
\end{figure}

To give a measure of the estimation error, we define
\begin{equation}\label{mse2}
ASE(\bA^*_{(1)})= \frac{1}{p}\sum_{i=1}^p{(\hat A^*_{1i}-A^*_{1i})^2},
\end{equation}
where $A^*_{1i}$ for $i=1,\ldots,p$ are the true coefficients in the first row of matrix $\bA^*$, and $\hat A^*_{1i}$ are their estimated values.
The box plots in figure \ref{figure6} summarize the results of the estimations from 500 replications of the model with $p=100$ (at the top) and $p=500$ (at the bottom). We report the average squared error computed by (\ref{mse2}) in three different cases: the classic Yule--Walker estimator of the VAR model $\hat\bA_{VAR}^*$ on the left, our estimator $\hat\bA_{SDPD}^*(\bW)$ proposed in (\ref{AW}) with the known spatial matrix in the middle, and our estimator $\hat\bA_{SDPD}^*(\hat\bW)$ proposed in (\ref{AWhat}) with the estimated spatial matrix on the right.

Figure \ref{figure6} shows interesting results. First, note that the classic estimator $\hat\bA_{VAR}^*$ cannot be applied when $T\leq p$, and this is a serious drawback of the classic VAR models. On the other hand, the \emph{stationary SDPD} model is equivalently used to represent the same process but it can always generate an estimation result for all values of $T$ and $p$ regardless of whether $\bW$ is known or unknown. Moreover, if we compare the box plots, we can note that both the median and the variability of the estimators $\hat\bA_{SDPD}^*(\bW)$ and $\hat\bA_{SDPD}^*(\hat\bW)$ are remarkably lower than those relative to the classic estimator $\hat\bA_{VAR}^*$ (when available) for all sample sizes $T$ and dimensions $p$. This deserves a further remark: while it is expected that the estimator $\hat\bA_{SDPD}^*(\bW)$ performs better than $\hat\bA_{VAR}^*$ (given that it exploits the knowledge of the true spatial matrix $\bW$), it is surprising to also see that the estimator $\hat\bA_{SDPD}^*(\hat\bW)$ outperforms the classic estimator $\hat\bA_{VAR}^*$, notwithstanding the fact that they function under the same conditions (only the time series $\by_t$ is observed and no spatial matrix is known). Of course, the ASE of the estimator $\hat\bA_{SDPD}^*(\hat\bW)$ slightly increases compared to that of the estimator $\hat\bA_{SDPD}^*(\hat\bW)$, but its variability remains more or less the same.




\appendix

\section{Appendix}
\begin{lemma}\label{lemma1}
Consider model (\ref{b1}) and its transformation (\ref{b1ter}), assuming a stationary process $\by_t$ with white noise errors $\bve_t$. If the matrix $\bSigma_0^{\varepsilon}=\mathop{var}(\bve_t)$ is diagonal (\emph{i.e.}, there is no cross-correlation in the error process), then the matrix $\bSigma_1=\mathop{cov}(\by_t, \by_{t-1})$ is symmetric.
\end{lemma}
\textbf{Proof.}
Assume for simplicity that the stationary process $\by_t$ has zero mean. By (\ref{b1ter}), we have
\begin{eqnarray}
\bz_t^{(0)}(\bz_{t-1}^{(0)})^T &=& D(\blambda_1)\bz_{t-1}^{(0)}(\bz_{t-1}^{(0)})^T + \bve_t(\bz_{t-1}^{(0)})^T \nonumber \\
E\left\{\bz_t^{(0)}(\bz_{t-1}^{(0)})^T\right\} &=& D(\blambda_1)E\left\{\bz_{t-1}^{(0)}(\bz_{t-1}^{(0)})^T\right\} + E\left\{\bve_t(\bz_{t-1}^{(0)})^T\right\} \nonumber \\
\bSigma^0_{1} &=& D(\blambda_1)\bSigma^0_0 \nonumber \\
\bSigma^0_{1}(\bSigma^0_0)^{-1} &=& D(\blambda_1). \label{ortogonali}
\end{eqnarray}
Now, under the assumption that $\bSigma_0^{\varepsilon}$ is diagonal (\emph{i.e.}, there is no spatial correlation in the error process), it is easy to show that $\bSigma_0^0$ is also diagonal. Therefore, by (\ref{ortogonali}) and the diagonality of matrix $D(\blambda_1)$, the matrix $\bSigma_1^0$ must also be diagonal.

So, defining $\bW^{(0)}=\left[\bI_p-D(\blambda_0)\bW\right]$ and remembering that $\bz_t^{(0)}=\bW^{(0)}\by_t$, we can write
\begin{eqnarray*}
\bSigma^0_{1} &=& \mathop{cov}(\bz_t^{(0)}, \bz_{t-1}^{(0)})=E\left\{\bW^0\by_t\by_{t-1}^T\bW^{0T}\right\}=\bW^0\bSigma_1\bW^{0T} \\
\bSigma_1 &=& (\bW^0)^{-1}\bSigma^0_{1}(\bW^{0T})^{-1} \equiv \left((\bW^0)^{-1}\bSigma^0_{1}(\bW^{0T})^{-1}\right)^T.
\end{eqnarray*}
All this implies that matrix $\bSigma_1$ is symmetric.
\hfill $\Box$\\

\begin{lemma}\label{lemma2}
Consider model (\ref{b1}) and its transformation (\ref{b1ter}). Under assumptions $A1$-$A4$, for fixed $p$ we have $\bw_i^T(\bSigma_1-\lambda_{1i}\bSigma_0)\bw_i\neq 0$ for all $i=1,\ldots,p$. Moreover, $\bw_i^T\bSigma_1\bw_i\neq 0$, $\bw_i^T\bSigma_0\bw_i\neq 0$ and $\lambda_{0i}\neq 0$.
\end{lemma}
\textbf{Proof.}
Without loss of generality, we assume that $\lambda_{0i}\neq 0$ and $\lambda_{1i}\neq 0$ for all $i=1,\ldots,p$. Given (\ref{vincolo}), we can write $(\be^T_i-\lambda_{0i}\bw^T_i)[\bSigma_1-\lambda_{1i}\bSigma_0]={\bf 0}$ so that
\begin{equation}\label{combinazione}
\be^T_i[\bSigma_1-\lambda_{1i}\bSigma_0]=-\lambda_{0i}\bw^T_i[\bSigma_1-\lambda_{1i}\bSigma_0],
\end{equation}
from which we can see that the $i$-th row of matrix $[\bSigma_1-\lambda_{1i}\bSigma_0]$ is equal to a linear combination of the other rows using the weights $\lambda_{0i}\bw_i$. Remember that $\bw_i$ has the $i$-th component equal to zero and $\bw_i\neq{\bf 0}$ by assumption $A4$. Therefore, matrix $[\bSigma_1-\lambda_{1i}\bSigma_0]$ is singular and has a rank of less than $p$. Without loss of generality, we assume that the coefficients $\lambda_{1i}$ are distinct for all $i$. If we post-multiply by $\bSigma_0^{-1}$, we get
\[
[\bSigma_1-\lambda_{1i}\bSigma_0]\bSigma_0^{-1}=\bSigma_1\bSigma_0^{-1}-\lambda_{1i}\bI_p,
\]
which has rank $p-1$ since matrix $\bSigma_1\bSigma_0^{-1}=\bA^*$ is diagonalizable by (\ref{diagonalize}), and $\lambda_{1i}$ is one of its eigenvalues (see Theorem 4.8 of \cite{Sch05}). Given that $\mathop{rank}(\bA\bB)\leq\min\left\{\mathop{rank}(\bA), \mathop{rank}(\bB)\right\}$, we conclude that matrix $[\bSigma_1-\lambda_{1i}\bSigma_0]$ has rank $p-1$; so it has $p-1$ linearly independent rows. By (\ref{combinazione}), we must conclude that the linear independent rows are those extracted by vector $\bw_i$, which identify a nonsingular submatrix extracted from $[\bSigma_1-\lambda_{1i}\bSigma_0]$. This sub-matrix is therefore definite positive or definite negative, and the first part of the Lemma is proved.

The last part of the Lemma is a direct consequence of the first one. In particular, note that the zero value is not admissible for the coefficient $\lambda_{0i}$ since it would be in contrast with the assumption $A4$. In fact, if $\lambda_{0i}=0$ for a given $i$, it means that there is not spatial correlation between the $i$-th location and the others, so that $\bw_i$ cannot be different from the null vector. Therefore, assumption $A4$ implies that $\lambda_{0i}\neq 0$ for all $i$ (but see also Remark 3).
\hfill $\Box$\\

\vspace{10pt}\noindent\textbf{Proof of Theorem \ref{theorem1}.}
Consider a given $i=1,\ldots,p$. To show point 1, from (\ref{vinc}) and assumption $A3$, we can write
\begin{equation}\label{vincolo_eigen}
\ba_i^T\bSigma_{1}\bSigma_0^{-1}=\lambda_{1i}\ba_i^T,
\end{equation}
where $\ba_i^T=(\be^T_i-\lambda_{0i}\bw^T_i)$. So, it is evident that $\lambda_{1i}$ is an eigenvalue of the matrix $\bSigma_{1}\bSigma_0^{-1}$, and $\ba_i$ is an eigenvector associated with $\lambda_{1i}$. Given that $\bw_i\neq{\bf 0}$ by assumption $A4$, the eigenvector $\ba_i$ is a continuous function of $\lambda_{0i}$. Now, remember that $\be_i$ has all zeroes except for a single instance of one in position $i$ while $\bw_i$ has a zero in position $i$. So, the vector $\ba_i$ has a one in position $i$, and this can be used in order to univocally identify the eigenvector $\ba_i$ among the eigenvectors associated with $\lambda_{1i}$.
Therefore, we can conclude that only one combination of eigenvalue/eigenvector satisfies (\ref{vincolo_eigen}) identifying a unique couple of values $(\lambda_{0i}^*,\lambda_{1i}^*)$.

Now, we analyse what happens if there are multiplicities in the vector $\blambda_1$. Under assumption $A2$, multiplicities of order $s<p$ in $\blambda_1$ are allowed since there are $p$ linearly independent eigenvectors in matrix $\left[\bI_p-D(\blambda_0)\bW\right]$. But if the multiplicity is of order $p$, which means that all the $\lambda_{1i}$s in vector $\blambda_1$ are the same, the equation system in (\ref{vinc}) would admit infinite solutions, and model (\ref{b1}) would not be identifiable (note that in such a case, the true model would be the classic SDPD of \cite{LeeYu10a} instead of the stationary SDPD proposed here).
However, the last case is excluded by assumption $A1$.

So, under assumptions $A1-A4$, the equation in (\ref{vinc}) actually admits only one solution. Denote such a solution with $(\lambda^*_{0i},\lambda^*_{1i})$. For point 2 of the Theorem, we show that this solution is also an extreme point of the following function:
\begin{equation}\label{FOC}
\lambda_{1i}=\mathop{cov}(z_{it}^{(0)}, z_{i,t-1}^{(0)})/\mathop{var}(z_{i,t-1}^{(0)})\equiv\lambda_{1i}(\lambda_{0i}).
\end{equation}
Now, we use the following arguments. We find the extreme points of $\lambda_{1i}$ as a function of $\lambda_{0i}$, and then, we check if one of these satisfies the constraint in (\ref{vinc}).

The first-order condition to find the extreme points of (\ref{FOC}) requires that
\[
\frac{\partial}{\partial\lambda_{0i}}\!\left\{\frac{\mathop{cov}(z_{it}^{(0)}, z_{i,t-1}^{(0)})}{\mathop{var}(z_{i,t-1}^{(0)})}\right\}=0.
\]
It can be equivalently expressed as follows:
\begin{eqnarray}
\frac{1}{\mathop{var}(z_{i,t-1}^{(0)})}\left[{\mathop{var}(z_{i,t-1}^{(0)})}\frac{\partial \mathop{cov}(z_{it}^{(0)}, z_{i,t-1}^{(0)})}{\partial\lambda_{0i}}-\mathop{cov}(z_{it}^{(0)}, z_{i,t-1}^{(0)})\frac{\partial\mathop{var}(z_{i,t-1}^{(0)})}{\partial\lambda_{0i}}\right] &=& 0 \nonumber\\
\frac{\partial \mathop{cov}(z_{it}^{(0)}, z_{i,t-1}^{(0)})}{\partial\lambda_{0i}}-\lambda_{1i}\frac{\partial\mathop{var}(z_{i,t-1}^{(0)})}{\partial\lambda_{0i}} &=& 0. \label{newcond}
\end{eqnarray}
Remembering that $z_{it}^{(0)}=(\be_i^T-\lambda_{0i}\bw^T_i)\by_t$, we can write
\begin{equation}\label{first_pol}
(a_{1i}b_{0i}-a_{0i}b_{1i})+(2a_{1i}c_{0i}-2a_{0i}c_{1i})\lambda_{0i}+(b_{1i}c_{0i}-b_{0i}c_{1i})\lambda_{0i}^2=0,
\end{equation}
where
\begin{eqnarray}
\mathop{cov}(z_{it}^{(0)}, z_{i,t-1}^{(0)}) &=&(\be_i^T-\lambda_{0i}\bw^T_i)\bSigma_1(\be_i-\lambda_{0i}\bw_i) \label{prima_riga}\\
&=& \be_i^T\bSigma_1\be_i-\be_i^T(\bSigma_1+\bSigma_1^T)\bw_i\lambda_{0i}+\bw^T_i\bSigma_1\bw_i\lambda_{0i}^2 = a_{1i} + b_{1i}\lambda_{0i} + c_{1i}\lambda_{0i}^2 \nonumber\\
\frac{\partial \mathop{cov}(z_{it}^{(0)}, z_{i,t-1}^{(0)})}{\partial\lambda_{0i}} &=& -(\be_i^T-\lambda_{0i}\bw^T_i)(\bSigma_1+\bSigma_1^T)\bw_i \label{dcov}\\
&=& -\be_i^T(\bSigma_1+\bSigma_1^T)\bw_i+2\bw^T_i\bSigma_1\bw_i\lambda_{0i} = b_{1i} + 2c_{1i}\lambda_{0i}\nonumber\\
\mathop{var}(z_{i,t-1}^{(0)}) &=& (\be_i^T-\lambda_{0i}\bw^T_i)\bSigma_0(\be_i-\lambda_{0i}\bw_i)\nonumber \\
&=& \be_i^T\bSigma_0\be_i-2\be_i^T\bSigma_0\bw_i\lambda_{0i}+\bw^T_i\bSigma_0\bw_i\lambda_{0i}^2 = a_{0i} + b_{0i}\lambda_{0i} + c_{0i}\lambda_{0i}^2\nonumber\\
\frac{\partial\mathop{var}(z_{i,t-1}^{(0)})}{\partial\lambda_{0i}} &=& -2(\be_i^T-\lambda_{0i}\bw^T_i)\bSigma_0\bw_i \label{dvar}\\
&=&-2\be_i^T\bSigma_0\bw_i+2\bw^T_i\bSigma_0\bw_i\lambda_{0i} = b_{0i} + 2c_{0i}\lambda_{0i}.\nonumber
\end{eqnarray}
So, condition (\ref{newcond}) has two solutions. We must check if one of these also satisfies the following equation system.
\begin{equation}\label{constraint}
(\be^T_i-\lambda_{0i}\bw^T_i)\bSigma_{1}-\lambda_{1i}(\be^T_i-\lambda_{0i}\bw^T_i)\bSigma_0={\bf 0}^T.
\end{equation}
Using (\ref{newcond}), (\ref{dcov}), (\ref{dvar}), and (\ref{constraint}), we can write
\begin{eqnarray}
\left.\frac{\partial\lambda_{1i}}{\partial\lambda_{0i}}\right|_{\lambda_{0i}=\lambda_{0i}^*} &=&
\left.\frac{\partial\mathop{cov}(z_{it}^{(0)}, z_{i,t-1}^{(0)})}{\partial\lambda_{0i}}\right|_{\lambda_{0i}=\lambda_{0i}^*}- \lambda_{1i}^*\left.\frac{\partial\mathop{var}(z_{i,t-1}^{(0)})}{\partial\lambda_{0i}}\right|_{\lambda_{0i}=\lambda_{0i}^*} \nonumber\\
&=& -(\be_i^T-\lambda_{0i}^*\bw^T_i)(\bSigma_1+\bSigma_1^T)\bw_i + \lambda_{1i}^*2(\be_i^T-\lambda_{0i}^*\bw^T_i)\bSigma_0\bw_i \nonumber\\
&=& -(\be_i^T-\lambda_{0i}^*\bw^T_i)(\bSigma_1+\bSigma_1^T)\bw_i + 2(\be^T_i-\lambda_{0i}^*\bw^T_i)\bSigma_{1}\bw_i  \nonumber \\
&=& -(\be_i^T-\lambda_{0i}^*\bw^T_i)(\bSigma_1^T-\bSigma_1)\bw_i\label{final},
\end{eqnarray}
which is equal to zero because $\bSigma_1$ is symmetric by Lemma \ref{lemma1}.
Therefore, for cross-uncorrelated errors $\bve_t$, the solution $(\lambda^*_{0i},\lambda^*_{1i})$ of (\ref{constraint}) also satisfies (\ref{newcond}). Thus, it identifies an extreme point of the correlation function in (\ref{FOC}).

\vspace{10pt}\noindent\textbf{Proof of Theorem \ref{theorem1bis}.}
Following the same arguments as in Theorem \ref{theorem1}, when $\bSigma_0^{\varepsilon}$ is a full matrix, we can write the following from (\ref{final}).
\begin{eqnarray*}
\left.\frac{\partial\mathop{cov}(z_{it}^{(0)}, z_{i,t-1}^{(0)})}{\partial\lambda_{0i}}\right|_{\lambda_{0i}=\lambda_{0i}^*}- \lambda_{1i}^*\left.\frac{\partial\mathop{var}(z_{i,t-1}^{(0)})}{\partial\lambda_{0i}}\right|_{\lambda_{0i}=\lambda_{0i}^*} &=& (\be_i^T-\lambda_{0i}^*\bw^T_i)(\bSigma_1^T-\bSigma_1)\bw_i,
\end{eqnarray*}
which is different from zero, given that matrix $\bSigma_1$ is not symmetric. Therefore, when the errors $\bve_t$ are cross-correlated, the solution $(\lambda^*_{0i},\lambda^*_{1i})$ does not represent an extreme point of the correlation function in (\ref{FOC}). Now, we define $a_{2i}=\be_i^T(\bSigma_1^T-\bSigma_1)\bw_i$, and we note that $-\bw_i^T(\bSigma_1^T-\bSigma_1)\bw_i=0$. So, remembering (\ref{prima_riga})-(\ref{dvar}), the previous equation can be written as
\begin{equation}\label{polinomio}
t_{0i} + t_{1i}\lambda_{i0} + t_{2i}\lambda_{i0}^2=0,
\end{equation}
where $t_{0i}=a_{0i}b_{1i}-a_{1i}b_{0i}+a_{0i}a_{2i}$, $t_{1i}=2a_{0i}c_{1i}-2a_{1i}c_{0i}+a_{2i}b_{0i}$, and $t_{2i}=b_{0i}c_{1i}-c_{0i}b_{1i}+a_{2i}c_{0i}$. (\ref{polinomio}) has two solutions, among which one is $(\lambda^*_{0i},\lambda^*_{1i})$.
Note that $a_{2i}=0$ when the matrix $\bSigma_1$ is symmetric, that is, under the assumptions of Theorem \ref{theorem1}. In such a case, (\ref{polinomio}) is equivalent to (\ref{first_pol}). Therefore, Theorem \ref{theorem1bis} includes Theorem \ref{theorem1} as a particular case.
\qqed\\

\vspace{10pt}\noindent\textbf{Proof of Theorem \ref{theorem3}.}
We show the first point of the theorem. For a given $i$, the second-order polynomial equation used to get the estimator is
\begin{eqnarray}
\hat t_{2i}\lambda_{0i}^2+\hat t_{1i}\lambda_{0i}+\hat t_{0i}=0 \label{eq1},
\end{eqnarray}
where the coefficients $\hat t_{0j}$s are functions of some entries of the matrices $\widehat\bSigma_0$ and $\widehat\bSigma_1$. Denote with $\bm_i$ a generic vector including a finite number (say, $K_i\leq 2p^2$) of elements of such estimated matrices, while $\bmu_i$ contains the corresponding true values from $\bSigma_0$ and $\bSigma_1$. It can be easily shown that
\[
\hat t_{0i}=g_0(\bm_i), \quad
\hat t_{1i}=g_1(\bm_i), \quad \hat t_{2i}=g_2(\bm_i),
\]
where the functions $g_j(\cdot)$s are differentiable.
By assumption $A3$, according to \cite{Han76}, we have
\[
\sqrt{T}(\bm_i-\bmu_i)\stackrel{d}{\longrightarrow} N({\bf 0},\bV_{0i}),
\]
where the elements of $\bV_i$ can be derived from \cite{Han76}.
It is also clear that $\hat\lambda_{0i}$ is a function of $\bm_i$, and we need to derive the gradient of such a function in order to prove the asymptotic normality of $\hat\lambda_{0i}$. To this aim, we use the theory of implicit functions. Let us indicate with $F(\bx,\lambda_0)$ the function in (\ref{eq1}), where $\bx$ is a vector of length $K_i$ denoting the vector of moments. Note that
\begin{eqnarray*}
\frac{\partial F}{\partial\lambda_0} &=& 2g_{2}(\bx)\lambda_0+g_{1}(\bx) \label{uno}\\
\frac{\partial F}{\partial\bx} &=& \left(\frac{\partial F}{\partial g_2}\frac{\partial g_2}{\partial\bx}+\frac{\partial F}{\partial g_1}\frac{\partial g_1}{\partial\bx}+\frac{\partial F}{\partial g_0}\frac{\partial g_0}{\partial\bx}\right)=\lambda_{0}^2 g'_2(\bx)+\lambda_{0} g'_1(\bx)+ g'_0(\bx).
\end{eqnarray*}
By Theorem \ref{theorem1bis}, $F(\bx,\lambda_0)=0$ for $\bx=\bmu_i$ and $\lambda_0=\lambda^*_{0i}$.
Moreover, $\left.\frac{\partial F}{\partial\lambda_0}\right|_{\bx=\bmu_i;\lambda_0=\lambda^*_{0i}}\neq 0$, since from (\ref{final}), we can write
\begin{eqnarray*}
\left.\frac{\partial F}{\partial\lambda_0}\right|_{\bx=\bmu_i;\lambda_{0i}=\lambda_{0i}^*} &=&
\left.\frac{\partial^2\mathop{cov}(z_{it}^{(0)}, z_{i,t-1}^{(0)})}{\partial\lambda_{0i}^2}\right|_{\lambda_{0i}=\lambda_{0i}^*}- \lambda_{1i}^*\left.\frac{\partial^2\mathop{var}(z_{i,t-1}^{(0)})}{\partial\lambda_{0i}^2}\right|_{\lambda_{0i}=\lambda_{0i}^*} \\
&& - \left.\left.\frac{\partial\lambda_{1i}}{\partial\lambda_{0i}}\right|_{\lambda_{0i}=\lambda_{0i}^*}\frac{\partial\mathop{var}(z_{i,t-1}^{(0)})}{\partial\lambda_{0i}}\right|_{\lambda_{0i}=\lambda_{0i}^*}\\
&=&2\bw_i^T(\bSigma_1-\lambda_{1i}^*\bSigma_0)\bw_i+2(\be_i^T-\lambda_{0i}^*\bw_i^T)\bSigma_0\bw_i\left.\frac{\partial\lambda_{1i}}{\partial\lambda_{0i}}\right|_{\bx=\bmu_i;\lambda_{0i}=\lambda_{0i}^*} \\
&=& S_1+S_2.
\end{eqnarray*}
At least one of these two terms is different from zero. In fact, note that $\left.\frac{\partial\lambda_{1i}}{\partial\lambda_{0i}}\right|_{\bx=\bmu_i;\lambda_{0i}=\lambda_{0i}^*}$ is zero if and only if the errors are cross-uncorrelated, by Theorem \ref{theorem1}. Therefore, in such a case, $S_2=0$, and we must assure that $S_1\neq 0$. The last is always guaranteed by assumption $A4$ and Lemma \ref{lemma2}.

As a result, we can conclude that $\left.\frac{\partial F}{\partial\lambda_0}\right|_{\bx=\bmu_i;\lambda_0=\lambda^*_{0i}}\neq 0$, and thanks to Dini's Theorem, there exists a function $f\in C^1$ such that $F(\bx,\lambda_0)=0$ if and only if $\lambda_{0}=f(\bx)$. Moreover,
\begin{equation}\label{grad}
\frac{\partial f(\bx)}{\partial\bx} = -\left(\frac{\partial F}{\partial\lambda_0}\right)^{-1}\frac{\partial F}{\partial\bx}=-\left({2g_2(\bx)\lambda_{0}+g_1(\bx)}\right)^{-1}\left(\lambda_{0}^2 g'_2(\bx)+\lambda_{0} g'_1(\bx)+ g'_0(\bx)\right).
\end{equation}


The estimator for $\lambda_{0i}$ is therefore $\hat\lambda_{0i}=f(\bm_i)$. Now, we use (\ref{grad}) to get the gradient of $f(\bx)$ at point $\bx=\bmu_i$, that is $\bD_i=\left.\frac{\partial f(\bx)}{\partial \bx}\right|_{\bx=\bmu_i}$.
Looking at the functions $g_j(\cdot)$ and $g'_j(\cdot)$ for $j=0,1,2$, it is evident that $\hat\lambda_{0i}$ depends on $\bm_i$ by means of the quantities $\bw_i^T\bSigma_0\bw_i$ and $\bw_i^T\bSigma_1\bw_i$. Therefore, the number of nonzero values in $\bD_i$ depends on the number of nonzero values in $\bw_i$. Given assumption $A4$, there is at least one value different from zero in vector $\bw_i$.
As a result, given that $f(\bmu_i)=\lambda_{0i}$, by a classical convergent result (section 3.3 in \cite{Ser80}), we find that $f(\bm_i)=\hat\lambda_{0i}$ is asymptotically normal.

Following the same argument as before, we can also prove the result in point 2 of the theorem, after noting that $\hat\lambda_{1i}=\rho(\hat\lambda_{0i})$. We can apply the \emph{Wold device} knowing that $\sqrt{T}(\hat\lambda_{0i}-\lambda_{0i}, \bm_i-\mu_i)^T$ is asymptotically normal.  The proof is now complete.
\qqed

\vspace{10pt}\noindent\textbf{Proof of Theorem \ref{theorem4}.}
By (\ref{grad}), the estimators $\hat\lambda_{0i}=f(\bm_i)$ and $\hat\lambda_{1i}=\rho(f(\bm_i))$ depend on the moment vector $\bm_i$ 
by means of the functions
\[
g_0(\bm_i)=\hat a_{0i}\hat b_{2i}-\hat a_{1i}\hat b_{0i}, \quad g_1(\bm_i)= 2\hat a_{0i}\hat c_{1i}- 2\hat a_{1i}\hat c_{0i}+\hat a_{2i}\hat b_{0i}, \quad g_2(\bm_i)=\hat b_{0i}\hat c_{1i}-\hat c_{0i}\hat b_{2i},
\]
where $\hat b_{2i}=-2\be_i^T\widehat\bSigma_1\bw_i$ is the estimated version of $b_{2i}=-2\be_i^T\bSigma_1\bw_i$ and the other terms can be derived from (\ref{prima_riga})-(\ref{dvar}). So,
the moments in $\bm_i$ are the estimated covariances from the $p\times p$ matrices $\hat\bSigma_0$ and $\hat\bSigma_1$, and the convergence rates of $\widehat\lambda_{0i}$ and $\widehat\lambda_{1i}$ derives from the (slowest of the) convergence rates of $\hat a_{ji}, \hat b_{ji}$ and $\hat c_{ji}$. 

We analyze first the component $\hat a_{ji}=\be_i\widehat\Sigma_j\be_i$ for given $i,j$. Note that it does not depend on $\bw_i$ and it involves only a single diagonal element of matrix $\hat\bSigma_j$, with $j=0,1$. By \cite{Han76} we can say that
\begin{equation}\label{ratea}
|\hat a_{ji}- a_{ji}|=|\be_i^T(\widehat\bSigma_j-\bSigma_j)\be_i|=O_p(T^{-1/2}) \qquad \forall j, \forall i {\rm \ and\ }\forall p.
\end{equation}

Now we analyze the components $\hat b_{ji}$ and $\hat c_{ji}$ for given $i,j$. Note that $\hat b_{ji}-b_{ji}\propto\bw_i^T(\widehat\bSigma_j-\bSigma_j)\be_i$ and $\hat c_{ji}-c_{ji}=\bw_i^T(\widehat\bSigma_j-\bSigma_j)\bw_i$. So, they are linear combinations of the elements in $\widehat\bSigma_j-\bSigma_j$. In order to analyze their asymptotic magnitude when $p\rightarrow\infty$, we need to analyze first the asymptotic behaviour of the vectors $\bw_i$, the coefficient $\lambda_{0i}$ and the matrices $\bSigma_j$ when $p\rightarrow\infty$. In particular, we need to check if the conditions for the identifiability of the model, stated in Theorem \ref{theorem1} and Lemma \ref{lemma2}, are still valid when $p\rightarrow\infty$.

By the (\ref{diagonalize}), the reduced form of model (\ref{b1}) is driven by
\begin{equation}\label{Astar}
\bA^*=\left[\bI_p-D(\blambda_0)\bW\right]^{-1}D(\blambda_1)\left[\bI_p-D(\blambda_0)\bW\right]=\bSigma_1\bSigma_0^{-1},
\end{equation}
where the last equality comes from the classic Yule-Walker equation system on model (\ref{b1bis}). 
So, the structure of $\bSigma_j$ for $j\in\{0;1\}$ depends on $\blambda_0$ and $\bW$, both of which may be influenced by $p\rightarrow\infty$. 
We may have different cases, among which the following cases (i)--(iii) of the theorem.

Consider the cases (i) and (ii) first, where the vectors $\bw_i$ are normalized by $L_r$ norm with $r=1,2$, respectively. These are typical setups in spatial models. Therefore, $\|\bw_i\|_r=1$ and $w_{ij}=O(p^{-1/r})$ for all $j\neq i=1,\ldots,p$. We can write
\begin{equation}\label{wi}
\bw_i=O(p^{-1/r})\bu_i
\end{equation}
where $\bu_i$ is a vector with $k_i$ ones and zeroes in the other positions (among which position $i$). 

Now, we let the spatial coefficients $\lambda_{0i}$ depend on the dimension $p$, following the idea in \cite{KelPru10}. 
If we assume that $\lambda_{0i}=O(p^{\gamma})$ for all $i$ and a given $\gamma$, then each one of the nonzero elements of matrix $D(\blambda_0)\bW$ is $O(p^{\gamma-1/r})$. So, by (\ref{Astar}), $\bSigma_1$ and $\bSigma_0$ tend to be diagonal for $p\rightarrow\infty$ if $\gamma<1/r$ and the non-diagonal elements tend to zero with rate $O(p^{\gamma-1/r})$.

Therefore, in order to guarantee the identifiability of the model under this setup, remember by Theorem \ref{theorem1} and Lemma \ref{lemma2} that we must have $c_{ji}=\bw_i^T\bSigma_j\bw_i\neq 0$ for $j=0,1$ and for all $p$. Given that $k_i=O(p)$ we have
\[
c_{ji}=\bw_i^T\bSigma_j\bw_i=p^{-2/r}\bu_i^T\bSigma_j\bu_i=p^{-2/r}k_i^2O(p^{\gamma-1/r})=O(p^{2+\gamma-3/r})
\]
so we need $2+\gamma-3/r\geq 0$, which means that we must have $\gamma\geq -1/2$ if $r=2$ or $\gamma\geq 1$ if $r=1$.

We can now derive the rates for $\hat c_{ji}$ and $\hat b_{ji}$, under the conditions above on $\gamma$ and $r$. We can refer to \cite{CheAlt13} to derive the convergence rate for $\widehat\bSigma_0$ when $p\rightarrow\infty$. In particular, following example 2.2 of \cite{CheAlt13}, we can use Theorem 2.1 and Corollary 2.2 case (iv), with $\alpha>1/2-1/q$ and $q>2$, to show that $E|\widehat\bSigma_0-\bSigma_0|^2_F=O(p^{2}T^{-1})$, where $|\cdot|_F$ is the Frobenius norm. By Schwartz inequality, this convergence rate can be extended also to the entries of matrix $\widehat\bSigma_1$. Now, we can write
\begin{eqnarray}
&&E(\hat c_{ji}-c_{ji})^2=E\!\left[\bw_i^T(\widehat\bSigma_j-\bSigma_j)\bw_i\right]^2=E\!\left[\sum_{r\neq i}\sum_{k\neq i}w_{ir}(\hat\sigma^{(j)}_{rk}-\sigma^{(j)}_{rk})w_{ik}\right]^2 \nonumber\\
&\leq&(\sup|\bw_i|)^4E\!\left\{\sum_{r\neq i}\sum_{k\neq i}(\hat\sigma^{(j)}_{rk}-\sigma^{(j)}_{rk})^2 +\sum_{r\neq i}\sum_{k\neq i}(\hat\sigma^{(j)}_{rk}-\sigma^{(j)}_{rk})\mathop{\sum_{s\neq i}\sum_{l\neq i}}_{s\neq r \cup l\neq k}(\hat\sigma^{(j)}_{sl}-\sigma^{(j)}_{sl})\right\} \nonumber\\
&=& R_1 + R_2, \label{r1r2}
\end{eqnarray}
where $R_1\leq(\sup|\bw_i|)^4 E|\widehat\bSigma_j-\bSigma_j|_F^2=O(p^{-4/r})O(p^2T^{-1})$, using the (\ref{wi}) and Theorem 2.1 of \cite{CheAlt13}. The term $R_2$ is of order smaller than (or equivalent to) $R_1$. In fact, notwithstanding $R_2$ is a sum of $p^4$ terms, most of them are deviations which are taken with signs and not in absolute value, so they tend to compensate each other and to vanish in the sum. The only terms in $R_2$ which do not compensate in the sum are those where $r=l\cap k=s$, because in such cases the pair of deviations tend to have the same sign and to be positively correlated, so that the product $(\hat\sigma^{(j)}_{rk}-\sigma^{(j)}_{rk})(\hat\sigma^{(j)}_{sl}-\sigma^{(j)}_{sl})$ has systematically positive sign. But the total number of such terms is less than $p^2$.

Following the same arguments, we can derive the convergence rates for $|\hat b_{ji}-b_{ji}|$. Finally, using Markov's inequality, we can transform the $O(\cdot)$ rates into $O_p(\cdot)$ rates. 

Summing all up, in case (i), where we consider $r=1$ and $\gamma=1$, we get the following convergence rates
\[
|\hat c_{ji}-c_{ji}|=O_p(p^{-1}T^{-1/2}), \qquad\qquad |\hat b_{ji}-b_{ji}|=O_p(p^{-1/2}T^{-1/2}) \qquad {\rm for\ }p\rightarrow\infty,
\]
so that, given the (\ref{ratea}), the final convergence rate for $|\widehat\lambda_{ji}-\lambda_{ji}|$ is $O_p(T^{-1/2})$.

In case (ii), where we have $r=2$ and $\gamma=0$, we get the following convergence rates
\[
|\hat c_{ji}-c_{ji}|=O_p(T^{-1/2}), \qquad\qquad |\hat b_{ji}-b_{ji}|=O_p(T^{-1/2}) \qquad {\rm for\ }p\rightarrow\infty,
\]
so that the final convergence rate for $|\widehat\lambda_{ji}-\lambda_{ji}|$ is $O_p(T^{-1/2})$. So, in both the cases, the dimension $p$ does not affect the rate. Note that the last case ($r=2$ and $\gamma=0$) has been considered in the simulation study.

In case (iii), the weights $\bw_i$ are generic (not normalized but bounded) vectors and $\lambda_{0i}=O(1)$ for all $i$ as $p\rightarrow\infty$. Note that in this case we have $\sup(|\bw_i|)=O(1)$. Therefore, the term $R_1$ in the (\ref{r1r2}) is $O_p(p^2T^{-1})$ by \cite{CheAlt13}, so that the final rate for $|\hat\lambda_{ji}-\lambda_{ji}|$  is $O(pT^{-1/2})$.
\qqed

\section*{Acknowledgments}
The author thanks professor Qiwei Yao and professor Francesco Giordano for helpful and stimulating discussions. This work has been partially supported by the PRIN research grant ``La previsione economica e finanziaria: il ruolo dell'informazione e la capacit\`{a} di modellare il cambiamento'' - prot. 2010J3LZEN005.

\newpage

\begin{table}[t]
\centering\footnotesize
\begin{tabular}{|c|cccc|cccc|} \hline
& \multicolumn{4}{c|}{$ASE(\widehat\blambda_0)$} & \multicolumn{4}{c|}{$ASE(\widehat\blambda_1)$} \\
$T$ & $p=10$ & $p=50$ & $p=100$ & $p=500$ & $p=10$ & $p=50$ & $p=100$ & $p=500$ \\ \hline
& \multicolumn{8}{c|}{$\bW_1$} \\ \hline
 50  &  0.3422  &  0.2955  &  0.243  &  0.247  &  0.0238  &  0.0246  &  0.022  &  0.0216  \\
 & (0.7591) & (0.3466) & (0.1452) & (0.0777) & (0.0162) & (0.0076) & (0.0043) & (0.0021) \\
 100  &  0.1902  &  0.1679  &  0.1533  &  0.1727  &  0.0111  &  0.0114  &  0.0101  &  0.0102  \\
 & (0.4842) & (0.1256) & (0.098) & (0.0699) & (0.0073) & (0.0033) & (0.002) & (9E-04) \\
 500  &  0.0347  &  0.0447  &  0.0497  &  0.0698  &  0.0019  &  0.0022  &  0.002  &  0.002  \\
 & (0.0645) & (0.034) & (0.043) & (0.0533) & (0.001) & (7E-04) & (4E-04) & (2E-04) \\
 1000  &  0.0173  &  0.0245  &  0.0264  &  0.0422  &  0.001  &  0.0011  &  0.001  &  0.001  \\
 & (0.0158) & (0.026) & (0.0219) & (0.0185) & (5E-04) & (3E-04) & (2E-04) & (1E-04) \\ \hline
& \multicolumn{8}{c|}{$\bW_2$} \\ \hline
 50  &  0.2425  &  0.4992  &  0.286  &  0.3467  &  0.0212  &  0.0236  &  0.0204  &  0.0221  \\
 & (0.3758) & (1.3098) & (0.1945) & (0.293) & (0.0129) & (0.0054) & (0.0033) & (0.0017) \\
 100  &  0.1152  &  0.3581  &  0.1814  &  0.2478  &  0.0092  &  0.0112  &  0.0095  &  0.0103  \\
 & (0.1796) & (0.3482) & (0.0868) & (0.173) & (0.005) & (0.0025) & (0.0015) & (8E-04) \\
 500  &  0.0134  &  0.1502  &  0.0716  &  0.1462  &  0.0017  &  0.0021  &  0.0017  &  0.0019  \\
 & (0.0094) & (0.1232) & (0.0645) & (0.6157) & (9E-04) & (4E-04) & (3E-04) & (2E-04) \\
 1000  &  0.0065  &  0.1125  &  0.0423  &  0.0939  &  8E-04  &  0.001  &  8E-04  &  9E-04  \\
 & (0.0078) & (0.1723) & (0.0312) & (0.0507) & (4E-04) & (3E-04) & (1E-04) & (1E-04) \\ \hline
& \multicolumn{8}{c|}{$\bW_3$} \\ \hline
 50  &  0.3712  &  0.4159  &  0.2413  &  0.3494  &  0.022  &  0.0199  &  0.0209  &  0.0206  \\
 & (0.5788) & (0.9416) & (0.1771) & (0.1505) & (0.0127) & (0.0042) & (0.0033) & (0.0014) \\
 100  &  0.2708  &  0.2134  &  0.1627  &  0.2386  &  0.0103  &  0.0094  &  0.0096  &  0.0096  \\
 & (1.0124) & (0.2002) & (0.1843) & (0.1391) & (0.0053) & (0.0021) & (0.0015) & (7E-04) \\
 500  &  0.0564  &  0.0902  &  0.0709  &  0.0992  &  0.002  &  0.0018  &  0.0017  &  0.0018  \\
 & (0.0733) & (0.0714) & (0.0692) & (0.0416) & (9E-04) & (4E-04) & (3E-04) & (1E-04) \\
 1000  &  0.0267  &  0.0752  &  0.0481  &  0.0672  &  0.001  &  9E-04  &  8E-04  &  9E-04  \\
 & (0.0305) & (0.0803) & (0.0486) & (0.0249) & (5E-04) & (2E-04) & (1E-04) & (1E-04) \\
\hline
\end{tabular}
\caption{\label{tabella1} \small Mean values (with standard deviations in brackets) of the average square error $ASE(\hat\blambda_{1})$ and $ASE(\hat\blambda_{0})$ for different sample sizes $T$, dimensions $p$, and spatial matrices $\bW$.}
\end{table}

\end{document}